\documentclass[12pt,a4paper,english]{article}

\usepackage[bindingoffset=0.3cm,textheight=22.5cm,hdivide={2.7cm,*,2.7cm}, vdivide={*,22cm,*}]{geometry}
\usepackage[dvips,bookmarksnumbered=true,breaklinks=true]{hyperref}

\usepackage{amsmath,amsfonts,amssymb,babel,slashed}

\usepackage[numbers,square,comma,sort&compress]{natbib}

\IfFileExists{dsfont.sty}
	{\usepackage{dsfont}
         \newcommand{\id}{\mathds{1}}}
	{\typeout{Package dsfont.sty was not found, using alternative macros.}
         \let\mathds=\mathbb
         \newcommand{\id}{\mbox{1 \kern-.59em {\rm l}}}}
\let\one=\id

\makeatletter

         \@addtoreset{equation}{section}
\makeatother

\let\startappendix=\appendix
\newcommand{\nocontentsline}[3]{}
\newcommand{\tocless}[3]{\bgroup\let\addcontentsline=\nocontentsline#1{#2}#3\egroup}

\newcommand{\Appendix}[1]{
  \refstepcounter{section}
  \section*{Appendix \thesection:\hspace*{1.5ex} #1}
  \addcontentsline{toc}{section}{Appendix \thesection}
}

\newcommand{\qed}{\nobreak \ifvmode \relax \else
      \ifdim\lastskip<1.5em \hskip-\lastskip
      \hskip1.5em plus0em minus0.5em \fi \nobreak
      \vrule height0.75em width0.5em depth0.25em\fi}

\newcommand{\be}{\begin{equation}}
\newcommand{\ee}{\end{equation}}
\newcommand{\eq}[1]{(\ref{#1})}

\def\nn{\nonumber}
\def\bea{\begin{eqnarray}}
\def\eea{\end{eqnarray}}
\def\obar{\overline}

\def\beqa{\begin{eqnarray}} 
\def\eeqa{\end{eqnarray}} 
\def\beq{\begin{equation}} 
\def\eeq{\end{equation}}

\def\Tr{{\rm Tr}}

\def\a{\alpha}          
\def\b{\beta}           
\def\c{\gamma}  
\def\d{\delta}    
\def\e{\epsilon}                
        
\def\g{\gamma} 

\def\k{\kappa}
\def\l{\lambda} \def\L{\Lambda}

\def\s{\sigma}  

\def\th{\theta}

\def\cA{{\cal A}}  \def\cC{{\cal C}}
 \def\cE{{\cal E}} 
 \def\cH{{\cal H}} \def\cI{{\cal I}}
\def\cJ{{\cal J}} \def\cK{{\cal K}} \def\cL{{\cal L}}
\def\cM{{\cal M}} \def\cN{{\cal N}} \def\cO{{\cal O}}
\def\cP{{\cal P}}  \def\cR{{\cal R}}
 \def\cT{{\cal T}}

\newcommand{\R}{\mathds{R}}

\newcommand{\mso}{\mathfrak{s}\mathfrak{o}}

\def\bit{\begin{itemize}}
\def\eit{\end{itemize}}

\def\({\left(}
\def\){\right)}

\def\d{\delta}

\def\pa{\partial} \def\del{\partial}

\newcommand{\tr}{\mbox{tr}}

\def\bcomment#1{}

\def\LNC{\Lambda_{\rm NC}}
\def\YM{{\rm YM}}

\newcommand{\co}[2]{[#1,#2]}						
\renewcommand{\a}{\alpha}
\renewcommand{\b}{\beta}
\renewcommand{\d}{\delta}

\renewcommand{\th}{\theta}

\renewcommand{\l}{\lambda}

\renewcommand{\L}{\Lambda}

 \allowdisplaybreaks[3]

\begin{document}

\renewcommand{\title}[1]{\vspace{10mm}\noindent{\Large{\bf
#1}}\vspace{8mm}} \newcommand{\authors}[1]{\noindent{\large
#1}\vspace{5mm}} \newcommand{\address}[1]{{\itshape #1\vspace{2mm}}}


\begin{flushright}
UWThPh-2012-33\\
\end{flushright}

\begin{center}

\title{ \Large The curvature of branes, currents and  gravity  \\[1ex]
 in matrix models}

\vskip 3mm
 
 \authors{Harold Steinacker}
 
 \vskip 3mm
 
 \address{ {\it Faculty of Physics, University of Vienna\\
  Boltzmanngasse 5, A-1090 Vienna, Austria  } }

\vskip 1.4cm

\textbf{Abstract}

\end{center}

\vskip 0.4cm

The curvature of brane solutions in Yang-Mills matrix models is expressed in terms of  
conserved currents associated with global symmetries of the  model. 
This implies a relation between the Ricci tensor and the energy-momentum tensor due to the basic 
matrix model action, without invoking an Einstein-Hilbert term. The coupling is governed by the extrinsic curvature
of the brane embedding, which arises naturally for compactified brane solutions. The effective gravity 
on the brane is thereby related to the compactification moduli, and protected from quantum corrections due to 
the relation with global symmetries.

\vskip 1.4cm

\newpage

\tableofcontents

\section{Introduction}
\label{sec:background}


Matrix models  provide remarkable candidates for a pre-geometric
 theory of fundamental interactions
including gravity. In particular, the IKKT model  \cite{Ishibashi:1996xs}
was proposed originally as a non-perturbative definition of IIB string theory, and the BFSS model  
(\cite{Banks:1996vh}, cf. \cite{deWit:1988ig})
as a non-perturbative definition of M-theory. 
It is well-known that the models  admit indeed brane solutions, 
consistent with IIB supergravity resp. 11-dimensional supergravity.
However they clearly go beyond supergravity, 
and should provide a non-perturbative quantum theory by integrating over the 
space of matrices. In particular the IKKT model is well suited for numerical simulations, 
and evidence was reported recently  \cite{Kim:2011cr} for the emergence of 3+1-dimensional space-time.
Therefore a better theoretical understanding of the dynamics of branes and their 
geometry and gravity is very important.

A systematic study of the effective geometry  of brane solutions in the matrix
model was undertaken in recent years \cite{Steinacker:2007dq,Steinacker:2008ri,Steinacker:2008ya,Steinacker:2010rh}. 
This led to a description of branes as
quantized symplectic submanifold embedded in $\R^{9,1}$, with effective metric 
$G^{ab} \sim \theta^{aa'}\theta^{bb'} g_{a'b'}$
 determined by the Poisson structure and the embedding metric $g$. 
 The dynamical metric $G^{ab}$ governs all matter and fields  on the branes,
 and must therefore be interpreted as  gravitational metric.
The relation with string theory or supergravity is seen by relating $\theta$ with the $B$ field, $g$ with the 
closed string (bulk) metric, and $G$ with the open string metric on the brane \cite{Seiberg:1999vs}.

Since the basic solutions of the model are  branes with a dynamical metric, one is  led to a picture
of brane-worlds. The mechanism for gravity is however not obvious, and there are several possible scenarios. 
One possibility is that quantization leads to an induced gravity action, which is however delicate
and leads to fine-tuning issues. 
Another  mechanism\footnote{A rather different gravitational interpretation of the IKKT equations of motion was proposed in 
\cite{Hanada:2005vr}, whose 
significance for the brane solutions is not clear. }
is holography, and indeed the bulk metric of 
supergravity seems to arise quantum mechanically from the brane description 
\cite{Maldacena:1997re,Kabat:1997sa,Ferrari:2012nw,Ishibashi:1996xs,Chepelev:1997av,Blaschke:2011qu}.
However, to obtain an acceptable 4-dimensional gravity the 10-dimensional bulk must be compactified in this scenario,
leading to a landscape of vacua with its inherent lack of predictivity \cite{Susskind:2003kw}.

On the other hand, these conventional pictures miss the basic fact that the  metric is not a 
fundamental degree of freedom  in the matrix model, but a derived quantity. This means that the 
geometrical equations of motion admit  solutions  of the brane embedding 
and its Poisson structure given by harmonic brane embeddings and  excitations
of $\theta^{ab}$. This suggests a different, ``emergent'' gravity mechanism, based on the basic matrix model action
rather than quantum effects. Indeed excitations of the 
Poisson structure lead to Ricci-flat perturbations \cite{Rivelles:2002ez} on flat $\R^4$ and certain self-dual 
geometries \cite{Yang:2006mn}. However on flat branes, matter does not seem to induce the  metric perturbations
 required for gravity. Remarkably, it {\em does} on branes with non-trivial extrinsic curvature 
as pointed out in  \cite{Steinacker:2009mp}, 
and indeed Newtonian gravity arises within harmonic brane excitations. 
A similar mechanism is realized on compactified brane solutions $\cM^4 \times \cK \subset \R^{10}$,
where the extrinsic curvature arises from the
compactification,  whose moduli become 4-dimensional gravitational modes \cite{Steinacker:2012ra}.
This mechanism is very interesting, but its analysis was limited to the linearized regime, and 
obscured by certain ``mixing'' terms whose significance remained unclear.

In this paper, we establish
new techniques and results which allow to efficiently compute the curvature of branes in the IKKT model. 
This provides significant new insights into the above mechanism for emergent gravity. 
We first establish a description of the geometry in terms of an over-complete frame, based on the 
currents associated with the global $SO(D)$ symmetry of the model. 
The curvature can then be computed using techniques from projective modules. 
We obtain an explicit and compact expression for a broad class of geometries 
including generalized almost-K\"ahler geometry, adapted to the case of Minkowski signature.
This class of geometries is argued to be sufficiently general and dynamically preferred by the model.
The currents are useful because their conservation law encodes the equations of motion of the 
brane, and moreover the energy-momentum tensor of matter acts as source for currents.
Using the results for the curvature, it follows that the energy-momentum tensor 
couples indeed to the Ricci tensor, albeit in an indirect way mediated by a tensor $\cP$ 
which also provides an additional vacuum contribution.
This coupling $\cP$ depends on the extrinsic curvature of the brane as well as the Poisson tensor,
hence on the  brane compactification.
Assuming that $\cP$ respects the effective 4-dimensional Lorentz symmetry, the Einstein equations 
should be recovered, up to vacuum contributions.
Although more work is required to clarify $\cP$ and its dynamics, this provides
strong support for the emergent gravity scenario on compactified 
branes  $\cM = \cM^4 \times \cK \subset \R^D$ in matrix models.

The relation with global symmetries and with non-commutative gauge theory 
make this mechanism for gravity very attractive for quantization, for the maximally
supersymmetric IKKT model. Since a compactification on fuzzy extra dimensions $\cK_N$ can be  
viewed as non-trivial vacuum in a $U(N) $ noncommutative $\cN=4$ SYM theory, the model is 
expected to be UV finite on such backgrounds. This is no longer the case for
more than 4 noncompact dimensions, which may explain the 
emergence of 3+1 non-compact dimensions as observed numerically in \cite{Kim:2011cr}.
Furthermore, since the currents are associated with global symmetries of the model, 
the gravitational degrees of freedom can be expected to remain massless at the quantum level, and
the mechanism should be protected from fine-tuning problems.

This paper is organized as follows. After reviewing the description of 
noncommutative branes in matrix models, 
we introduce the generalized frame formalism
for the embedding geometry in section \ref{sec:embedding-frame}  based on the  currents. 
This is extended to the 
effective geometry in section \ref{sec:effective-frame}, 
where a special class of geometries is introduced. The Ricci tensor for the effective geometry
is then computed in section \ref{sec:Ricci}. We also discuss the equation of motion for 
the Poisson structure via  energy-momentum conservation in section \ref{sec:e-m-cons}, 
and include a section on flux stabilization.
The inhomogeneous current conservation law is derived in section \ref{sec:perturbations},
and some technical details are elaborated in the appendices.

\section{Matrix models and their geometry} 
\label{sec:matrixmodels-intro}

We briefly collect the essential ingredients of the matrix model framework
and its effective geometry, referring 
to the  review \cite{Steinacker:2010rh} for more details.

\subsection{The IKKT model}
\label{sec:basic}

The starting point is given by a matrix model of Yang-Mills type, 
\begin{align}
S &=-\frac {\L_0^4}4\Tr\(\co{X^A}{X^B}\co{X^C}{X^D}\eta_{AC}\eta_{BD}\,  +  2 \obar\Psi \gamma_A[X^A,\Psi] \)   
 = S_{\YM} + S_\Psi
\label{S-YM}
\end{align}
where the $X^A$ are Hermitian matrices, i.e. operators acting on a separable Hilbert space $\mathcal{H}$. The
indices of the matrices run from $0$ to $D-1$, and
will be raised or lowered with  the invariant tensor $\eta_{AB}$ of $SO(D-1,1)$.
We also introduce a parameter  $\L_0$ of dimension $[L]^{-1}$, so that the  $X^A$ have dimension length.
We focus on the maximally supersymmetric IKKT or IIB model \cite{Ishibashi:1996xs} with $D=10$, 
which is best suited for quantization. It is obtained from the $\cN=1$ 
$U(N)$ SYM on $\R^{10}$ dimensionally reduced to a point,
and taking $N \to \infty$.
Then $\Psi$ is a matrix-valued Majorana Weyl spinor of $SO(9,1)$.
The model enjoys the fundamental gauge symmetry 
\be
X^A \to U^{-1} X^A U\,,\qquad  \Psi \to U^{-1} \Psi U\,,\qquad    U \in U(\cH)\,
\label{gauge-inv}
\ee
as well as the 10-dimensional Poincar\'e symmetry
\be
\begin{array}{lllll}
X^A \to \L(g)^A_B X^b\,,\quad  &\Psi_\a \to \tilde \pi(g)_\a^\b \Psi_\b\,,\quad  & g \in \widetilde{SO}(9,1), &  \\[1ex]
X^A \to X^A + c^A \one\,,\quad & \quad & c^A \in \R^{10}\,\quad 
\end{array}
\label{poincare-inv}
\ee
and a $\cN=2$ matrix supersymmetry \cite{Ishibashi:1996xs}.
The tilde indicates the corresponding spin group.
We  define the matrix Laplacian as
\begin{align}
  \Box \Phi :=  [X_B,[X^B,\Phi]]
\end{align}
for any matrix $\Phi \in \cL(\cH)$.
Then  the equations of motion of the model take the following form
\be
 \Box X^A \,=\, [X_B,[X^B,X^A]] =  0 ,
\label{eom-IKKT}
\ee
 assuming  $\Psi = 0$. Analogous statements hold more generally to matrix models of Yang-Mills type, 
 with Euclidean or Minkowski signature.

\subsection{Noncommutative branes and their geometry}

Now we focus on matrix configurations which describe embedded
noncommutative (NC) branes. This means that 
the $X^A$ can be interpreted as quantized embedding functions \cite{Steinacker:2010rh}
\be
X^A \sim x^A: \quad \mathcal{M}^{2n}\hookrightarrow \R^{10} 
\ee 
of a $2n$- dimensional submanifold of $\R^{10}$. More precisely,
there should be some quantization map $\cI: \cC(\cM)  \to  \cA\subset L(\cH)$
which maps classical functions on $\cM$ to a noncommutative (matrix) algebra of functions, 
such that commutators can be interpreted as quantized Poisson brackets. In the
semi-classical limit indicated by $\sim$,  matrices are identified with functions via $\cI$,
and commutators are replaced by Poisson brackets; for a more extensive introduction
see e.g. \cite{Steinacker:2010rh,Steinacker:2011ix}.
One can then locally choose $2n$ independent coordinate functions
$x^a,\ a = 1,...,2n$ among the $x^A$, and their commutators 
\begin{align}
\co{X^a}{X^b} \ &\sim \  i \{x^a,x^b\} \ = \ i \th^{ab}(x)\,
\end{align}
encode a quantized Poisson structure on $(\cM^{2n},\theta^{ab})$.
These $\theta^{ab}$ have dimension $[L^2]$ and set a typical scale of noncommutativity $\LNC^{-2}$.
We will assume that $\th^{ab}$ is non-degenerate\footnote{If the Poisson structure is degenerate, 
then the fluctuations propagate only
along the symplectic leaves.}, so that the 
inverse matrix $\th^{-1}_{ab}$ defines a symplectic form 
on $\mathcal{M}^{2n}\subset\R^{10}$. This submanifold 
is equipped with the induced metric
\begin{align}
g_{ab}(x)=\pa_a x^A \pa_b x_A\,
\label{eq:def-induced-metric}
\end{align}
which is the pull-back of $\eta_{AB}$.  However, this is {\em not} the effective metric on $\cM$. 
To understand the effective metric and gravity, we need to consider matter on the brane $\cM$. 
Bosonic matter or fields arise from 
nonabelian fluctuations of the matrices around a stack $X^A \otimes \one_n$
of coinciding branes, while fermionic matter arises
from $\Psi$ in \eq{S-YM}.
It turns out that  in the semi-classical limit, the  effective action for such fields is
governed by a universal effective metric $G^{ab}$. 
It can be obtained most easily by considering the action of an additional 
scalar field $\phi$ coupled to the matrix model in a gauge-invariant way, with action
\begin{align}
 S[\phi] &= -\frac{\L_0^4}2\Tr\, [X_A,\phi][X^A,\phi] 
\sim \frac{\L_0^4}{2(2\pi)^n} \int d^{2n} x \sqrt{|\theta^{-1}|} \th^{aa'}\th^{bb'}g_{a'b'}\, \del_a\phi \del_b \phi \nn\\
 &= \frac{\L_0^4}{2(2\pi)^n}  \int d^{2n} x\sqrt{| G|} \, G^{ab} \del_a\phi \del_b \phi .
 \label{action-scalar-geom}
\end{align}
Therefore  the effective metric is given by \cite{Steinacker:2008ri}
\begin{align}
 G^{ab}&= e^{-\sigma} \g^{ab}\,,  \qquad \g^{ab}=  \th^{aa'}\th^{bb'}g_{a'b'}\, \nn\\
  e^{-\s}&=\Big(\frac{\det{\th^{-1}_{ab}}}{\det{ G_{ab}}}\Big)^{\frac 1{2}}\,  
 = \Big(\frac{\det{\th^{-1}_{ab}}}{\det{g_{ab}}}\Big)^{\frac 1{2(n-1)}}  . 
\label{eff-metric}
\end{align}
It is useful to consider 
the  conformally equivalent  metric\footnote{More abstractly, this can be stated as 
$(\a,\b)_\g = (i_\a\theta,i_\b \theta)_g$ where $\theta = \frac 12\theta^{ab}\del_a\wedge\del_b$.}
$\g^{ab}$ which satisfies
\begin{align}
\sqrt{|\theta^{-1}|} \g^{ab} &= \sqrt{|G|} \,G^{ab}  .
\end{align}
The effective metric $G^{ab}$ is  encoded in the matrix Laplace operator, which 
can be seen from the following result  \cite{Steinacker:2010rh} for the semi-classical limit 
\begin{align}
 {\bf\Box} \Phi &= [X_A,[X^A,\Phi]]  
\ \sim\  -e^\s \Box_{G}\, \phi 
\end{align}
acting on scalar fields $\Phi \sim \phi$.
In particular, the matrix equations of motion \eq{eom-IKKT} take the  simple form
$\Box X^A \,\sim\, -e^\s \Box_{G} x^A = 0$.
This means that the embedding functions $x^A \sim X^A$ are harmonic functions with respect to $G$.
Furthermore, the bosonic part of the matrix model action \eq{S-YM} can be written in the semi-classical limit as follows
\begin{align}
 S_{\YM}\ \sim \ \frac{\L_0^4}{4(2\pi)^{2n}}\int d^{2n} x \sqrt{|\theta^{-1}|}\, \g^{ab}g_{ab} .
\label{action-semiclass}
\end{align}

\paragraph{Compactified brane solutions.}

Of particular interest here are branes with compactified extra dimensions
\be
\cM^{2n}  = \cM^4 \times \cK \quad \subset \ \R^D
\label{compactification}
\ee
where the extrinsic curvature is predominantly due to $\cK \subset \R^{D}$, while the embedding
of $\cM^4$ is approximately flat.
Such solutions including $\cK = T^2$ and $\cK = S^3 \times S^1$ 
have been given recently \cite{Steinacker:2011wb},
where $\cK$ is rotating along $\cM^4$ and stabilized by angular momentum. This
is possible because of  "split noncommutativity``, where the
Poisson structure  relates the compact space $\cM^4$ with the non-compact space $\cK$,
\be
\theta^{\mu i} = \{x^\mu, y^i\} \neq 0 \ .
\label{split-NC}
\ee 
Here $x^\mu$ are coordinates on $\cM^4$ and $y^i$ are coordinates on $\cK$.
As pointed out in \cite{Steinacker:2012ra}, such a structure relates the perturbations of $\cK$ to 
perturbations of the effective  metric on $\cM^4$, and thereby links the Ricci tensor to the 
energy-momentum tensor. This leads to a novel 
mechanism for 4-dimensional  gravity. 
The aim of this paper is to understand better this mechanism, by computing the intrinsic
curvature  on the brane in the presence of matter.

\section{Currents and geometry}

To compute the curvature  of $\cM$ directly from the metric $\g_{ab}$ is complicated and not illuminating. 
We will develop a suitable generalized frame formalism,
which allows to express the curvature efficiently in terms of the $SO(D)$  conserved currents.
This will be the key to a better understanding of the effective  gravity on the branes.

\subsection{Currents and conservation laws}

The matrix model  \eq{S-YM} is invariant under the $SO(D)$ resp. $SO(1,D-1)$ 
symmetries\footnote{When we write $SO(D)$ usually  $SO(1,D-1)$ will also be understood.}
\begin{align}
\d X^A = (\l^\a)^A_B X^B \, 
\end{align}
for $\l^\a \in \mso(D)$. Setting $\Psi = 0$ for now,
 they lead to conserved currents
 in complete analogy to quantum field theory,
\begin{align}
[X^A, \tilde J^{\a}_A] = 0, \qquad 
\tilde J^{\a}_A &=  \frac 12 \l^\a_{CD}\{X^C,[X_A, X^D]\} 
\label{current-cons-matr}
\end{align}
with anti-symmetric $\l^\a_{AB} \in \mso(D)$. This can be verified directly using the equation of 
motion \eq{eom-IKKT}, or more conceptually via a matrix version of the Noether theorem, 
as elaborated in  appendix \ref{sec:currents}.
In the semi-classical limit, this reduces to 
\begin{align}
\nabla^a J_{a}^{\a} \   &= 0   \qquad  J_a^{\a}  = x^A \l_{AB}^\a \del_a x^B\nn\\[1ex]
\tilde J^{\a}_A  \ &\sim  \  i \theta^{ab}\del_a x_A J_b^{\a} ,  
\label{currents-semiclass}
\end{align}
where $\nabla$ is the Levi-Civita connection corresponding to the effective metric $G$.
The conservation law in the presence of matter will be discussed in section \ref{sec:cons-current-matter}.
These conservation laws  completely  capture the equation of motion for the modes which 
preserve $S^{D-1} \subset \R^D$. Due to their origin from global symmetries, 
these conservation laws 
are expected to be protected from quantum corrections, as usual in quantum field theory.
This makes them well suited to describe the geometry of the model and its dynamics.

\subsection{Generalized embedding frame}
\label{sec:embedding-frame}

The above currents are  naturally viewed as one-forms in the 
cotangent bundle $T^*\cM$,
\begin{align}
  J^\a  =  x^A \l_{AB}^\a d x^B   = J_a^\a d\xi^a , \qquad  J_a^\a = x \l^\a \del_a x
\end{align}
(dropping the $\R^D$ indices)
where $\l^\a = \l^\a_{AB} \in \mso(D)$, and $\xi^a$ denotes any local coordinates on $\cM$.
It is useful to supplement them with the ''radial`` current corresponding to $\l^0 = \one$,
\begin{align}
J_a^0 = x \l^0 \del_a x, \qquad J^0 = x_A d x^A  = J_a^0 d\xi^a = r dr
\end{align}
where
\begin{align}
  r^2 = x_A x^A 
\end{align}
is the invariant radius on $\R^D$.
The basic observation underlying this paper is that these currents provide a generalized, over-complete frame 
for the metric $g$. More precisely, define the one-forms
\begin{align}
 \theta^{\a}_a =  r^{-1} J^\a_a , \qquad \theta^\a = r^{-1} J^\a .
\end{align}
Then the following identity\footnote{No equation of motion or current conservation is needed here.} holds
\begin{align}
 g_{ab}  = \k_{\a\b}\, \theta^{\a}_a \theta^{\b}_b = r^{-2}  \k_{\a\b} J_a^\a J_b^\b .
\end{align}
where
\begin{align}
\k^{\a\b} &= \begin{pmatrix}
               1 & 0 \\
               0 & -\frac 12 tr \l^\a \l^\b
             \end{pmatrix}  
\end{align}       
is the Killing form of $\mso(D)$ resp. $\mso(1,D-1)$ supplemented by $\l^0$.
This can be seen using the identity
\begin{align}         
 \k_{\a\b}\l^\a_{AB} \l^\b_{CD} &= \eta_{AC}\, \eta_{BD} - \eta_{AD} \, \eta_{BC} + \eta_{AB}\, \eta_{CD} ,
\label{lambda-identity}
\end{align}
which is easy to check for the basis of $\l^\a$  given by 
\begin{align}
 \l^{(AB)}_{CD} = \d^A_C\d^B_D - \d^A_D\d^B_C , \qquad A<B \quad\mbox{and}\ \  \l^{0}_{CD} = \eta_{CD}
\end{align}
where $\k^{\a\b}= \d^{\a\b}$ in the Euclidean case.
Correspondingly, 
\begin{align}
   P^{\a\b} =  \theta^\a_a \theta^\b_b g^{ab}, \qquad  P^\a_{\ \b} \theta^\b = \theta^\a
\end{align}
 is a projector on the cotangent bundle $T^*\cM$; the frame indices 
will always be raised and lowered with $\k_{\a\b}$, e.g. $P^\a_{\ \b} = P^{\a\g}\k_{\b\g}$.
The projector on the normal bundle is then given by
\begin{align}
P_N^{\b\g} = \k^{\b\g} - P^{\b\g} ,  \qquad  (P_N)^\a_{\ \b} \theta^\b = 0 .
\label{normal-projector}
\end{align}
Since the frame $\theta^\a$ is (over-) complete,
any one-form on $\cM$ can be written  as
$v = \sum_\a \theta^\a v_\a = \sum_\a \theta^\a P_\a^{\ \b} v_\b$. This expansion
is unique if we impose that $P v = v$.
In other words, the space of one-forms on $\cM$ can be 
identified with the projective module\footnote{The label $\cE_g$ indicates that the frame $\theta^\a$
encodes the metric $g$, to distinguish it from $\cE_\g$ introduced below. Module means that 
the elements can be multiplied (from the right, most naturally) with functions $f\in\cA$.} 
$\Omega^1(\cM) \cong \cE_g := P\cA^N$, where $\cA = \cC(\cM)$. 
This construction turns out to be very useful to compute the curvature.

Furthermore, the following identity is shown in appendix \ref{sec:deriv-currents}:
\begin{align}
 d \theta^{\b} P_{\b}^{\ \a} &=  -\theta^{\b}  \omega_{\b}^{\ \a} 
   \label{dtheta-result}
\end{align}
where
\begin{align}
 \omega^{\b\a} &= r^{-1}(\theta^\b P^{0\a} - P^{0\b} \theta^\a) = -  \omega^{\a\b} .
 \label{omega-AS}
\end{align}
which satisfies $P\omega = \omega = \omega P$.

\subsection{Connection and curvature}
\label{sec:conn-curv-I}

The above realization of the cotangent bundle as as projective module $T^*\cM \cong\cE_g$ 
is useful, because it provides  a canonical (``Grassmann'')
connection
\begin{align}
 \nabla_g = P \circ d: \quad \cE_g  \to \cE_g \otimes_\cA \Omega^1(\cA) .
\end{align}
The curvature of this  connection is defined by
\begin{align}
 \cR[g] = \nabla_g^2 = P dP dP ;
\end{align}
for an introduction to these concepts see e.g. \cite{Landi:1997sh}.
The last identity follows using $\theta = P\theta$ from
\begin{align}
\nabla_g^2 \psi &= P d(P d(P\psi)) 
= (P dP dP) \psi .
\end{align}
Under gauge transformations $v \to \L v \in \cE$ which commute with $P$, 
the connection transforms as $\nabla_g \to \nabla_g' = \L\nabla_g\L^{-1} = \nabla_g +   P \L^{-1}d\L$, 
and the curvature $\nabla_g^2 = PdPdP$ transforms in the adjoint.
Furthermore,  $\nabla_g$ is  compatible with the inner product
on $\cE_g$ which arises 
from  $\k^{\a\b}$ restricted to $\cE = P\cA^N$,
\begin{align}
 d(v,w)_g &= (\nabla_g v,w)_g + (v,\nabla_g w)_g ,
\end{align}
where 
\begin{align}
 (v,w)_g &=  v w^\dagger = v_\a \k^{\a\b} w_\b = v_\a P^{\a\b} w_\b  = v_a g^{ab} w_b
\end{align}
because $Pw = w, \ Pv = v$. This is the usual metric compatibility condition. 
The gauge transformations are compatible with this 
inner product if $\L^\dagger  = \L^{-1}$, where
\begin{align}
 (v^\dagger)^\a = \k^{\a\b}\, v_\b \ , \qquad (\L^\dagger)^{\ \a}_{\b} = \k_{\d\b}\L^{\ \d}_{\g}\k^{\a\g} 
\end{align}
etc.
Finally, the torsion $T:\cE \cong \Omega^1(\cA) \to \Omega^2(\cA)$ is defined as\footnote{The present definitions entail
$m \circ \nabla(v f) = \nabla(v)f + v\wedge df$ so that gauge covariance holds.}
\begin{align}
  T(v f) = (d + m\circ\nabla)(v f) =  T(v) f 
\end{align}
where $m(\theta\otimes \a) = \theta\wedge \a$.

Let us compute the Grassmann connection explicitly.
The one-form $\theta^\b$ is represented by
\begin{align}
 \theta^\b = \theta^\a P_\a^{\ \b}  \ \cong (P_\a^{\ \b}) \quad \in\ \cE = P \cA^N
\end{align}
and its covariant derivative is
\begin{align}
 \nabla_g \theta^\b =  \theta^\a \otimes(P_\a^{\ \g} d P_\g^{\ \b}) .
\end{align}
 Therefore 
$P d P$ is the  connection coefficient.
The torsion is given by
\begin{align}
 T^\g &= d \theta^\g + m(\nabla_g \theta^\g) = d \theta^\g + \theta^\a \wedge P_\a^{\ \b} d P_\b^{\ \g}  \nn\\
 &=  d (\theta^\b P_\b^{\ \g}) + \theta^\b\wedge d P_\b^{\ \g} \nn\\
 &= d\theta^\b P_\b^{\ \g}  = - \theta^\b \omega_\b^{\ \g}  
\label{torsion-universal-P}
\end{align}
using\footnote{Note that $\omega$ is not the spin connection, and the curvature is 
not given by $d\omega + \omega\omega$.} the result \eq{dtheta-result}.
More general connections can be defined as $\nabla \to \nabla +  A$
where $A = A^\a_{\ \b} \in\Omega^1(\cA)$ 
such that $PA = A = A P$, which is metric compatible if $A^\dagger = - A$. 
In particular, the torsion vanishes if we choose $A = \omega$,  
\begin{align}
0 =  T^\a[g] = T^\a + \theta^\b \wedge \omega_\b^{\ \a}  = 0 .
\label{torsion-vanish}
\end{align}
This is compatible with the metric since $\omega^\dagger = - \omega$ \eq{omega-AS}, and therefore
 $\nabla[g] = \nabla_g + \omega$  is the Levi-Civita
connection on $\cM$ for $g$. 

The curvature  2-form $\cR^\a_{\ \b}$ is a linear map on $\cE_g \cong T^*\cM$,
which can be written in standard tensorial form using the frame. 
The Grassmann curvature can  be evaluated  easily noting that $\theta P_N = 0$
along with $dP = - dP_N$ and $P^\dagger = P$:
\begin{align}
 \cR[g] &= \nabla[g]^2 = P dP dP  \nn\\
\cR_{ac}[g] &= \theta_a \cR\theta_c^\dagger =  \theta_a dP_N dP_N\theta_c^\dagger = d\theta_a P_Nd\theta_c^\dagger    \nn\\[1ex]
  &= r^{-2} \del_e J^\b_a (P_N)_{\b\g} \del_f J_c^\g\, dx^e dx^f
\label{univ-curv-explicit}
\end{align} 
using $d\theta P_N = r^{-1} dJ P_N $ in the last step.

\paragraph{Normal embedding coordinates (NEC).}
To evaluate this, we first choose suitable coordinates  at any given point $p\in\cM$:
using the rotation symmetry of the model, we can assume that $T_p\cM \cong (\R^{2n},0,...,0)$.
We can then choose the first $2n$ matrix components $x^a$ as local  coordinates, 
 denoted as  ``normal embedding coordinates'' NEC. It follows that
 $\del x_A \del\del x^A = 0$ at $p\in\cM$, which implies $\del|_p g_{ab} = 0$.
 Therefore these are indeed normal coordinates for $g$ in the Riemannian sense, so that 
we are essentially using $\nabla[g]$. 
 We can furthermore assume $p = (0,...,0,r_0)$ after a suitable translation, so that
 \begin{align}
 x_A\del x^A|_p = r\del r|_p = J^0|_p =  0.
 \end{align}
 Now consider the following  tensors
\begin{align}
 T_{ab}^\a &= \del_a x \l^\a \del_b x  \nn\\
K_{ab}^\a &=  x \l^\a  \nabla_a[g] \del_b x = x \l^\a  \del_a \del_b x|_p =  K_{ba}^\a \ ,
\label{T-K-def}
\end{align}
in particular $T_{ab}^0 = g_{ab}$.
Clearly $K_{ab}^\a$ characterizes the exterior curvature of $\cM \subset \R^{D}$.
Then
\begin{align}
  \nabla_{a}[g] J_b^\a  &=  T_{ab}^\a +   K_{ab}^\a = \del_{a}J_b^\a|_p  \ ,
 \label{J-T-K-id}
\end{align}
We note that  $T_{ab}^\a= -  T_{ba}^\a$ for $\a\neq 0$. 
It follows using \eq{lambda-identity}  that 
 \begin{align}
 K_{ea}^\a \k_{\a\b} J^{\b}_d &= 0 =  K_{ea}^\a P_{\a}^{\ \b}\nn\\
T_{ea}^\a \k_{\a\b} J^{\b}_d &= J^0_e g_{ad} - J^0_a g_{ed} + J^0_d g_{ea} \nn\\
 T_{ea}^\a P_{\a}^{\ \b}|_p &= 0 =  T_{ea}^\a P_{\a\b} T_{fc}^\b |_p \ .
\label{K-T-orthogonal}
\end{align}
Therefore\footnote{Note that $(\nabla_a J_b) \,J_c^\dagger$ does {\em not}vanish  identically.
This is the reason why the final result \eq{embedded-curv-explicit} contains additional radial contributions $\nabla  J^0$, and 
is not obtained trivially by re-shuffling $\nabla J \nabla J$. } 
$K^\a$ and $T^\a$ live in the normal bundle at $p$, and
dropping the contributions of $J^0$ at $p$
the Grassmann curvature 2-form for the metric $g$ is 
\be
\fbox{$
 \cR[g]_{ac} = r^{-2} \nabla_e J_a \, \nabla_f J_c^\dagger\, dx^e dx^f .
$}
\label{R-G-full}
\ee
where $\nabla = \nabla[g]$.
The  point is that the Grassmann curvature tensor can be expressed in terms of the 
$\mso(D)$ currents of the matrix model. 
We  now obtain the Riemann tensor for $g$ via
\begin{align}
 R[g] &= \nabla[g]^2 = P dP dP +  P (d \omega + \omega \omega)P \nn\\
R_{ac}[g] &= \cR_{ac}[g] + \theta_a(d\omega + \omega\omega)\theta_c^\dagger 
 \equiv  \cR_{ac}[g] +  \cR_{ac}[\omega].
\label{full-R}
\end{align}
Here $\omega$ takes care of the radial contributions which are not captured by the 
$\mso(D)$ currents, as computed in appendix \ref{sec:curv-rad}. Together with the above we obtain
\be
\fbox{$
 R[g]_{ac} =  r^{-2}\big(\nabla_e J_a \, \nabla_f J_c^\dagger\, - g_{ae} \nabla_f J^0_c   \, -  g_{cf} \nabla_e J_a^0  \,\big)\,  dx^edx^f
  \label{embedded-curv-explicit}
$} 
\ee
where $\nabla = \nabla[g]$, dropping again  contributions of $J^0|_p = 0$  and
recalling  $\nabla_a J^0_b = g_{ab} + K^0_{ab} = \frac 12 \nabla_a\del_b r^2$.
This is the key result, which will be extended to the effective metric $\g$ in the next section.
As a check, we proceed by decomposing $\nabla_a J_b^\a = T_{ab}^\a + K_{ab}^\a$ and using
\begin{align}
r \nabla_a\del_b r =  \frac 1{2}\nabla_a\del_b r^2 
 = \del_a x_A\del_b x^A +  x_A \nabla_a\del_b x^A 
 = g_{ab} +  x_A \nabla_a\del_b x^A 
 \label{deldelr-id}
\end{align}
along with the identity \eq{lambda-identity} to obtain
\begin{align}
  T_{ea}^\a \k_{\a\b} T_{fc}^\b &=  (\del_e x\l^\a\del_a x)(\del_f x\l_\a\del_c x) 
 = g_{ef} g_{ac} - g_{ec} g_{af} +   g_{ae} g_{cf}  \nn\\
 K_{ea}^\a \k_{\a\b} T_{fc}^\b  &= (x\l^\a\nabla_e\del_a x)(\del_f x\l_\a\del_c x) 
  = K^0_{ea}  g_{fc} \nn\\
  T_{ea}^\a  \k_{\a\b}  K_{fc}^\b  &=(\del_e x\l_\a\del_a x) (x\l^\a\nabla_f\del_c x)
   = g_{ea}  K^0_{fc}   \nn\\
K_{ea} P_N K_{fc} &=  K_{ea}^\a \k_{\a\b} K_{fc}^\b 
 = r^2 \nabla_e\del_a x_A \nabla_f\del_c x^A .
 \label{TK-contractions}
\end{align}
Since we assumed NEC, 
the coordinate-invariant form is obtained by replacing $\del_a \to \nabla_a[g]$.
We thus recover the usual Gauss-Codazzi theorem\footnote{This can  be obtained 
quickly using the projective module defined by the over-complete frame $\theta^A = d x^A$.}
for the Riemann curvature tensor on $\cM\subset \R^D$,
\begin{align}
 R[g]_{ac} &=  d\del_a x_A d\del_c x^A  
 = \frac 12(\nabla_e\del_a x_A \nabla_f\del_c x^A - \nabla_f\del_a x_A \nabla_e\del_c x^A) dx^e dx^f 
\end{align}

\subsection{Effective frame}
\label{sec:effective-frame}

We now want to develop a similar machinery for the effective metric $\g_{ab}$ on $\cM$.
This metric is encoded in the following (over-complete) tangent frame associated to the currents,
\begin{align}
 V^\a &= x \l^\a \{x,.\} = J^\a_a \theta^{ab} \del_b \qquad \in T\cM  \nn\\
 V^\a V^\b \k_{\a\b} &= r^2 \g^{ab}\del_a\otimes \del_b
\end{align}
including $\a=0$ as before. Here $\{.,.\}$ is the Poisson bracket on $\cM$, which arises from the non-commutative
nature of the brane.
However to compute the curvature, it is more natural to use the corresponding frame of one-forms, defined as usual
by lowering the index with the effective metric $\g$. Thus
\begin{align}
 \Theta^{\a} &=  \Theta_a^{\a}  d\xi^a, \qquad
 \Theta_a^{\a} =  r^{-1} V^{\a,b} \g_{ba}  =  \theta_b^{\a} \cJ^b_{\ a}
 \label{eff-frame}
\end{align}
 where
\begin{align}
  \cJ^c_{\ a} =  \theta^{cb} \g_{ba} =  \theta^{-1}_{ab} g^{bc} .
\end{align}
Then the effective metric can be written as
\begin{align}
\g_{ab} &= \kappa_{\a\b}\Theta^{\a}_a\Theta^{\b}_b = g_{a'b'} \cJ^{a'}_{\ a} \cJ^{b'}_{\ b} 
  = -g_{ac}\, (\cJ^2)^c_{\ b} 
 \label{eff-meetric-J}
\end{align}
and the tangential projector can  be expressed in various ways
 \begin{align}
  P^{\a\b} &= \Theta^{\a}_a\Theta^{\b}_b \g^{ab} 
   =  \theta^{\a}_c \theta^{\b}_d  \cJ^c_{\ a} \cJ^d_{\ b} \g^{ab}   
    = \theta^{\a}_a\theta^{\b}_b g^{ab} 
    = - \Theta^{\a}_a  \theta^{ae} \theta^{\b}_e  \nn\\
  P^{\a\b} \Theta^{\b} &= \Theta^{\a}, \qquad P^{\a\b} \theta^{\b} = \theta^{\a} .
  \label{P-forms}
\end{align}
Note that $P$ coincides with the projector defined in the previous section; this is evident due to the 
relation \eq{eff-frame} between the frames. 
The symplectic form $\Omega$ on $\cM$ is then given by
\begin{align}
 \Theta_a \theta_b^\dagger  &= \theta_c \cJ_{\ a}^{c} \theta_b^\dagger
  = \theta^{-1}_{ab} , \qquad 
 \Theta \theta^\dagger = \Omega .
\label{Tt-id}
\end{align}
A cotangent vector can now be written in the two bases as $v = \theta^\a v_\a = \Theta^\a  v_\a'$ 
with $P v = v, \ P v' = v'$. 
This gives two different identifications of $T^*\cM$ with  projective modules $\cE_g$ resp. $\cE_\g$.
We can determine the transformation $\L v' = v$
between the two frames explicitly, such that
\begin{align}
 \Theta^{\a} = \theta^\b \L_\b^{\ \a}
\end{align}
and therefore
\begin{align}
  \g_{ab} &=  \Theta_a\Theta_b^\dagger =  \theta_a \L \L^{\dagger} \theta^\dagger_b \ .
\label{g-gamma-L-rel}
\end{align}
This $\L$ is of course not unique. A nice invertible $\L$ which satisfies this requirement is given by
\begin{align}
 \L^{\a\b} &= P_N^{\a\b}  + \theta^\a_a \theta^\b_b \L^{ab}_{(AS)} \nn\\
  &= \L_{(S)} + \L_{(AS)}
\label{L-decomp}
\end{align}
where 
\begin{align}
\L^{ad}_{(AS)} &=  -\cJ^a_{\ c} g^{cd} = g^{ae} g^{dc} \,\theta^{-1}_{ec}  = - \L^{da}_{(AS)}
\label{L-explicit} \\
 \L_{(AS)}^\dagger &= - \L_{(AS)} 
\label{LAS-antiherm}
\end{align}
is anti-symmetric resp. anti-hermitian.
It satisfies
\begin{align}
 P \L = P \L_{(AS)} = \L_{(AS)}, \qquad P_N \L = P_N \ .
\end{align}
The inverse is given explicitly by
\begin{align}
 \L^{-1} = P_N + \theta_a \theta_b \theta^{ab} \ .
 \label{Lambda-inv}
\end{align}
We will accordingly define $\L^\a_{\ \b} = \L^{\a\b'} \k_{\b'\b}$ etc.
Note that the Poisson structure is encoded in $\L$, while the embedding is encoded in  $P$. 
Now consider the Grassmann connection on the  projective module $\cE_\g$, given by
\begin{align}
 \nabla_\g v &= \Theta^\a P_\a^{\ \b} d v_\b' \ = \theta \L P d (\L^{-1} v)  \nn\\
  \nabla_\g &=  \L \nabla_g \L^{-1}
\end{align}
Therefore $\nabla_\g$ is related to $\nabla_g$ 
via the (in general {\em non-orthogonal}) transformation $\L$. 
This is so because $\nabla_\g$ is compatible with the metric $\g$ encoded in
$(v, w)_\g := v'_\a  w'_\a =  v'_\a P^{\a\b} w'_\b$, while $\nabla_g$ is compatible with $g$.
The curvature $\nabla_\g^2$ acts on $\cE_\g \cong T^*\cM$  as follows
\begin{align}
 \nabla_\g^2 v &= \Theta^\b \cR^\a_\b[\g]  v_\a' = \theta  \L \cR[\g] \L^{-1} v  \nn\\
\nabla_g^2 v &= \theta \cR[g] v 
\end{align}
reflecting the fact that the  connections $\nabla_\g$ and $\nabla_g$ are  related by $\L$.
As in the previous section,
the coordinate form of the (Grassmann) curvature tensor can be obtained using the frame $\Theta^\a$
\begin{align}
\cR_{ab}[\g] &=  d\Theta_a P_N  d\Theta_b^\dagger  
 = \Theta_a dP_N dP_N \Theta_b^\dagger \nn\\
 &= \cR_{a'b'}[g] \cJ^{a'}_{\ a} \cJ^{b'}_{\ b} 
   =  \theta_a \L dP_N dP_N \L^\dagger \theta_b^\dagger   .
\label{R-gamma-expand}
\end{align}
As explained before,
the metric (Levi-Civita) connection corresponding to $\g$ is given by
\begin{align}
 \nabla[\g] = \nabla_\g + A[\g]
\end{align}
if $A = - A^\dagger$ is  such that the torsion vanishes,
\begin{align}
 T[\g] &= T_\g + \Theta A[\g] = 0 
\end{align}
To determine $A$, we compute
\begin{align}
  T_\g &= d \Theta + m(\nabla_\g \Theta) = d\Theta P \nn\\
 &=  d(\theta\L) P
 =  -\theta d\L P+ d\theta P\L\nn\\
  &= \Theta (d\L^{-1} \L P -\L^{-1}\omega\L)
\label{torsion-universal-P-eff}
\end{align}
using $d\theta P =- \theta \omega$ \eq{dtheta-result} and $P\omega = \omega$. 
Therefore the torsion $ T[\g]$ vanishes for
\begin{align}
A^{\a}_{\ \b}[\g] &= -P d\L^{-1} \L P  + P\L^{-1}\omega\L P  + \Theta^\a B^{(\a)}_\b \nn\\
&= P \L^{-1} d \L P + \L^{-1}\omega\L  + \Theta^\a B^{(\a)}_\b  
\label{A-gamma}
\end{align}
where $B_\b^{(\a)}$ is arbitrary 
(since $\Theta^\a \k_{\a\b}\Theta^\b = 0$). This is metric compatible if $A$ is anti-hermitian,
\begin{align}
 A^\dagger = -A \ .
\end{align}
The second term is always anti-hermitian due to \eq{LAS-antiherm}, \eq{omega-AS} and
 $\L^{-1}\omega\L = \L^{-1}_{(AS)}\omega\L_{(AS)}$.
In particular, the Grassmann connection is  torsion-free if $\L$ is unitary, which
is evident since then the metrics $g$ and $\g$ coincide \eq{g-gamma-L-rel}.

\paragraph{Conformal rescaling.}

Now consider the effective metric $G^{ab} = e^{-\s} \g^{ab}$ \eq{eff-metric}. The above construction
can easily be generalized by introducing a suitably rescaled frame 
\begin{align}
\tilde \Theta^\a_a &= e^{-\s/2 } \Theta^\a_a = \theta^{\a}_b \tilde\cJ^b_{\ a} = \theta^{\b}_a \,\tilde\L_\b^{\ \a}  \nn\\[1ex]
\tilde\cJ^b_{\ a} &= e^{-\s/2 } \cJ^b_{\ a} , \qquad \tilde\L = P_N + e^{-\s/2 } \L_{(AS)}
\end{align}
such that
\begin{align}
 \tilde\Theta^\a_a \tilde\Theta^\b_b \k_{\a\b} &= G_{ab} .
\end{align}
This leaves the projector $P$ unchanged.
However this kind of rescaling is more appropriate after compactification, and we will 
largely work with $\g^{ab}$ in this paper.

\subsection{Special geometry}

In  general, we cannot give an explicit form for the  $B^{(\a)}_\b$ required for the Levi-Civita connection.
We therefore restrict ourselves to a certain  class of preferred geometries.
More specifically, we consider geometries with 
\begin{align}
 \nabla[g] Q \equiv P dQ P = 0
\end{align}
where
\begin{align}
Q &:= \L\L^\dagger -\one   = - (\L_{(AS)}^2 + P)  
= -  \theta_a (\cJ^2 +\d)^a_{\ b} g^{bc} \theta^\dagger_{c} = \L^\dagger\L -\one \nn\\
 &= P Q = Q P \ .
\label{Q-const}
\end{align}
$Q$ measures the deviation from $\cJ$ being an almost-complex structure, 
in particular $Q=0$ for almost-K\"ahler geometries (in the Euclidean case). Together with \eq{eff-meetric-J}
and  $\nabla Q = \theta \nabla \cJ^2 g \theta^\dagger$ 
this implies  $\nabla[g] \g = \nabla[g] (g \cJ^2) =0$, so that  this condition is equivalent to 
\begin{align}
 \nabla \cJ^2 = 0 , \qquad \nabla[g] \equiv \nabla[\g] \equiv \nabla[G] \equiv \nabla .
\end{align}
The last equality follows from $\del \det\cJ^2 = 0$ together with \eq{eff-metric}.
This means that the connections on $\cM$ defined by $\g$ and $g$ and $G$ are equivalent, which is very reasonable.
Now $P dQ P = 0$ implies
\begin{align}
0 &= P d\L \L^\dagger P + P \L d\L^\dagger P  \nn\\
0 &= P\L^{-1} d\L P  + P(\L^{-1}d\L)^\dagger P 
\label{dL-L-spec-geom}
\end{align}
so that 
\begin{align}
 A_\L := P\L^{-1} d\L P  = \ \L_{(AS)}^{-1} d \L_{(AS)} \ = - A_\L^\dagger \ ,
\end{align}
using $0 = P dP P \equiv \nabla_g P$, and the Levi-Civita connection $\nabla[\g]$ 
is obtained for $B_\b^{(\a)}=0$.
Note that we do {\em not} require $\cJ^2 = -1$, which is impossible 
in the Minkowski case due to the inequivalent causal structures of $g$ and $\g$.
However $\nabla \cJ^2=0$ is  compatible with a Minkowski signature, and milder\footnote{which in turn is 
milder than e.g. the K\"ahler condition since $\cJ^2 \neq -1$.} than $\nabla \cJ =0$.
Typically $\cJ^2$  defines an (integrable) decomposition of $T\cM$ into rank 2 sub-bundles.
Moreover, it is not hard to see that 
the equations of motion for the Poisson structure $\theta^{ab}$
derived from the  bosonic action
\begin{align}
 S_{\rm YM} \sim \int d^{2n}\xi\, \sqrt{|\theta^{-1}|}\, \g^{ab} g_{ab} = - \int \Omega^{\wedge n}\,  \tr \cJ^{-2} 
\end{align}
are always satisfied if $\nabla \cJ^2 = 0$; this will become clear in section \ref{sec:e-m-cons}.
Moreover, geometries with $\nabla \cJ^2 = 0$ are not only solutions but are expected to be preferred
``ground state'' solutions for the Poisson structure.
This is true at least for 4-dimensional Euclidean branes, where the bosonic action is 
positive definite and takes its minimum if and only $\cJ^2 = -\d$ i.e. $Q=0$ \cite{Steinacker:2010rh}.

We therefore expect that $\nabla \cJ^2 = 0$ will always hold at least asymptotically.
However in general, $\nabla \cJ^2 = 0$ might not always be compatible with a given $g$, and
matter might lead to short-range perturbations of $\cJ^2$ or $\theta^{ab}$. 
As observed by Rivelles \cite{Rivelles:2002ez},
such perturbations are in fact Ricci-flat at least on $\R^4$.
Thus we expect 
that the Poisson structure is adjusted dynamically such that $\nabla \cJ^2 \approx 0$, and possible
deviations from $\nabla \cJ^2 = 0$ are  suppressed for long distances and 
could  be treated perturbatively. 
Special geometry should be even less restrictive in the presence of 
compactified extra dimensions, an compatible with all physically relevant
4-dimensional effective geometries.
The dynamics of $\cJ$ will be studied  in section \ref{sec:e-m-cons}.

\subsection{Curvature  and effective gravity}
\label{sec:Ricci}

Let us therefore assume special geometries with $\nabla Q  = 0$. 
Then the Levi-Civita connection  $\nabla$ is  given by 
\begin{align}
 \nabla = \nabla_\g + A , \qquad A \ = \ \L_{(AS)}^{-1} d \L_{(AS)}  + \omega
\label{nabla-gamma-explicit}
\end{align}
using \eq{L-decomp}.
After some algebra, we obtain the following expression for
the Riemann curvature for $\g$ (see appendix \ref{sec:spec-geom-curv}) using \eq{R-gamma-expand},
 \begin{align}
   R_{ab}[\g] &=  \theta_a (d P_Nd P_N +d\omega + \omega\omega) \L \Theta_b^\dagger  \nn\\[1ex]
      &= - R_{ab'}[g]\,  \cJ^{2 b'}_{\ \ b}  
      = - (\cR_{ab'}[g]\, + \cR_{ab'}[\omega])\, \cJ^{2 b'}_{\ \ b}  
   \label{R-gamma-Jsquare}
 \end{align}
 recalling that $\L_{(AS)}^\dagger = - \L_{(AS)}$.
 This also follows from\footnote{Note that the 
 non-trivial perturbations of $\g$  on $\R^4_\theta$ due 
 to fluctuations of the Poisson structure discussed e.g. in \cite{Rivelles:2002ez,Steinacker:2008ri} are not compatible 
with the assumption of special geometry, so there is no contradiction.} 
 $R^a_{\ b}[\g] \equiv R^a_{\ b}[g]$, since $\nabla[\g] = \nabla[g]$ for special geometries.
On the other hand, it follows from \eq{nabla-gamma-explicit} that $dA_\L + A_\L A_\L = 0$, so that 
the Riemannian curvature for $\g$ can be obtained from  the  Grassmann curvature 
via 
\begin{align}
 R_{ab}[\g] &= \cR_{ab}[\g] + \Theta_a \L^{-1} (d\omega + \omega \omega)\L \Theta_b^\dagger \   \nn\\
  &= \cR_{a'b'}[g]\cJ^{a'}_{\ a} \cJ^{b'}_{\ b} - \theta_a (d\omega + \omega \omega) \theta_{b'}^\dagger \cJ^{2b'}_{\ b} 
  \label{R-gamma-full}
\end{align}
using  \eq{R-gamma-expand} in the second line.
These are explicit and compact expressions for the effective curvature, which 
together with the expression \eq{R-G-full} for $\cR_{ab}[g]$ in terms of the currents
constitutes a  main result of this paper. 
Comparing the  two results \eq{R-gamma-full} and \eq{R-gamma-Jsquare} implies
$\cR_{a'b'}[g]\cJ^{a'}_{\ a} \cJ^{-1 b'}_{\ b}  = - \cR_{ab}[g]$, and noting that
$\cR[\omega]$  satisfies the standard symmetries of the 
Riemann tensor (e.g. using the explicit form \eq{omega-curv-explicit}) we have
\begin{align}
 \cR_{a'b';cd}[g]\cJ^{a'}_{\ a} \cJ^{-1 b'}_{\ b}  = - \cR_{ab;cd}[g]
  = \cR_{ab;c'd'}[g]\cJ^{c'}_{\ c} \cJ^{-1 d'}_{\ d} \ .
 \label{R-J-flip}
\end{align}
Now we compute the Ricci tensor from \eq{R-gamma-Jsquare}:
\begin{align}
 {\rm Ric}_{ac}[\g] &= \g^{bd} R_{ab;cd}[\g] = g^{bd} \cR_{ab;cd}[g] + g^{bd} \cR_{ab;cd}[\omega]  \
 = {\rm Ric}_{ac}[g] \ .
\label{ricci-eff-decomp}
\end{align}
Consider the two terms separately. 
For the first term, we use the relation \eq{R-J-flip} as follows
\begin{align}
 g^{bd} \cR_{ab;cd}[g] &=  g^{bd} \cR_{a'b';c'd'}[g]\cJ^{a'}_{\ a} \cJ^{-1 b'}_{\ b}\cJ^{c'}_{\ c} \cJ^{-1 d'}_{\ d}
  = \g^{bd} \cR_{a'b;c'd}[g]\cJ^{a'}_{\ a} \cJ^{c'}_{\ c} \, .
\end{align}
Now we can  use 
the explicit form of $\cR[g]$  in terms of the currents is given by \eq{R-G-full},
and together with \eq{TK-contractions} and \eq{deldelr-id} we obtain in NEC
\begin{align}
\nabla_a J_b^0 &= g_{ab} + K^0_{ab} = r \nabla_a \del_b r \nn\\
 \nabla^d J_d \, T_{ac}^\dagger &= 
  \g^{bd}(T + K)_{bd} T_{ac}^\dagger = \g^{bd} r \nabla_b\del_dr \, g_{ac}=  (\g^{bd}\nabla_b J_d^0)\, g_{ac}   \nn\\
 \g^{bd}\nabla_d J_b  \nabla_c J^\dagger_{a} &= \del_d(\g^{db} J_b)  \nabla_c J^\dagger_{a} 
  = \nabla^d J_d \, K_{ac}^\dagger +  (\g^{bd}\nabla_b J_d^0)\, g_{ac}  \nn\\
 &= \g^{bd} K_{db} \, K_{ac}^\dagger +  \g^{bd} K^0_{db}\, g_{ac} +  (\g^{bd}g_{bd})\nabla_a J^0_c\,   \nn\\
   \g^{bd}\nabla_c J_b\nabla_d J^\dagger_{a} &= \g^{bd}(T_{bc} + K_{bc})(T_{da}^\dagger + K_{da}^\dagger) \ .
\end{align}
using $J^0|_p = 0$.
To proceed, we  assume a compactified brane of the form $\cM^4\times \cK \subset \R^D$
where $\cK$ has a {\em small} radius of scale $r_K$, much smaller than any scale $r_\cM$ associated with the 
non-compact part.
Then the dominant terms are those arising from the 
extrinsic curvature on $\cK$, which is $K_{ab} K_{cd} \sim r_k^{-2}$. Therefore we  only keep the terms
quadratic in $K_{ab}^\a$ from now on and drop the rest, so that
\begin{align}
 \g^{bd}\nabla_d J_b  \nabla_c J^\dagger_{a} &= \nabla^d J_d \, K_{ac}^\dagger \ + \cO(\frac{r_\cK}{r_\cM})  \nn\\
  &= - \L_0^{-4}\, T_{cd}\,\Pi^{cd}_{ef}\, \theta^{ee'}\theta^{ff'} \, K_{e'f'}  K_{ac}^\dagger   \ + \cO(\frac{r_\cK}{r_\cM}) \nn\\
   \g^{bd}\nabla_c J_b\nabla_d J^\dagger_{a} &= \g^{bd} K_{bc} K_{da}^\dagger \ + \cO(\frac{r_\cK}{r_\cM}) \ 
\end{align}
using  current conservation  \eq{cons-law-matter}.
Furthermore, the contributions  \eq{R-omega-contract} from $g^{ae}\, \cR_{ac;ef}[\omega]$ are 
negligible in the same approximation.
Then the Ricci tensor for the effective metric is obtained from  \eq{ricci-eff-decomp} as
\begin{align}
 {\rm Ric}_{ac}[\g] &= r^{-2}\big(-\L_0^{-4}\, T_{cd}\,\Pi^{cd}_{ef}\, \theta^{ee'}\theta^{ff'} \, K_{e'f'}  K_{a'c'}^\dagger
   - \g^{bd}K_{bc'} K_{a'd}^\dagger\big) \cJ^{a'}_{\ a} \cJ^{c'}_{\ c} \nn\\
   &= r^{-2}\big(\L_0^{-4}\, T_{cd}\,\Pi^{cd}_{ef}\, \theta^{ee'}\theta^{ff'} \, K_{e'f'}  K_{a'c'}^\dagger
   + \g^{bd}K_{bc'} K_{a'd}^\dagger\big) \cJ^{2c'}_{\ c}
\end{align}
where the second form follows directly from \eq{R-gamma-Jsquare}.
The first line becomes more appealing (and more appropriate for the reduction to 4 dimensions, as explained below) 
in upper-component notation. Using also $\del e^\s = 0$, we obtain
a compact expression for the Ricci tensor provided $\nabla \cJ^2 = 0$,
\be
\fbox{$
e^{2\s} {\rm Ric}^{ac}[G] = {\rm Ric}^{ac}[\g] =  - T_{b'd'}\, \Pi^{b'd'}_{bd} \ \cP^{bd;ac} - \L_0^{4}\, g_{bd} \cP^{ab;cd} 
\  + \cO(\frac{r_\cK}{r_\cM})\ ,
$}
\label{Ricci-gamma-J-NC}
\ee
refining\footnote{The sign appears to be inconsistent with 
\cite{Steinacker:2012ra}.}  the previous results in \cite{Steinacker:2012ra}.
However to fully understand the effective gravity on $\cM$ we need to understand also the  response of
the second term 
$g_{bd}\cP^{ab;cd}$ to matter, which might contain an additional hidden coupling to $T_{ab}$.
Therefore this equation does not allow to draw immediate physical conclusions.
Nevertheless, the message is that the Ricci tensor is related to the energy-momentum tensor of matter,
{\em without} invoking an Einstein-Hilbert-type action or quantum effects. 
The coupling of matter to the Ricci-tensor is mediated  by the tensor
\begin{align}
 \cP^{cd;ab} &= r^{-2} \L_0^{-4}\,\theta^{cc'}\theta^{dd'} K_{c'd'} K_{a'b'}^\dagger\,\theta^{aa'}\theta^{bb'} 
  =   \L_0^{-4}\, \theta^{cc'}\theta^{dd'}\theta^{aa'}\theta^{bb'} \, \del_{c'}\del_{d'} x^A \del_{a'}\del_{b'} x_A      \nn\\
 \Pi^{cd}_{ab} &= \d^{cd}_{ab} - \frac1{2(n-1)} \g_{ab}\g^{cd}
 \label{pi-def}
\end{align}
which is determined by the extrinsic curvature  of the embedding $\cM = \cM^4 \times \cK\subset \R^D$
and the Poisson tensor $\theta^{ab}$. The second form of $\cP$ follows from \eq{TK-contractions}.
Without extrinsic curvature, $\cP$ would vanish, and not even 
Newtonian gravity would arise\footnote{Of course other mechanisms are conceivable such as induced gravity or holography.
However, then the usual fine-tuning problems would arise.}.
The last terms subsume the  ``mixing terms''  
which remained  mysterious in \cite{Steinacker:2012ra}.

The expression  \eq{Ricci-gamma-J-NC} should be a suitable starting point to study 
the effective gravity on branes, which 
will be pursued elsewhere. However we emphasize several  points here.
First, the  extrinsic curvature is necessarily large on 
the compact extra dimensions $\cK$,
which is transmitted to the {\em non-compact} space $\cM^4$ by 
the Poisson tensor  as in \eq{split-NC}. In this way, the compactification moduli of $\cK$ can play the role of
gravitational degrees of freedom for $\cM^4$, and their origin in the spontaneously broken 
global symmetries of the matrix model implies that they remain massless\footnote{This is not
the case for the radial modes, which were discussed in \cite{Steinacker:2012ra}. These are in 
fact assumed to be massive here due to the flux on $\cK$.}.
Such Poisson tensors $\theta^{\mu i}$ which relate the compact 
with the non-compact space
naturally arise on compactified brane solutions in matrix models,
dubbed split non-commutativity \cite{Steinacker:2011wb}.
For example, a spherical compactification $\cK = S^2\subset \R^6$ would lead to
$K_{ij} K_{kl}^\dagger = r_\cK^{-2} \d_{ij}\d_{kl}$, which is too simple to provide full 
Einstein gravity. However the $\cK$ typically 
has to rotate  along $\cM^4$ in order to be a solution \cite{Steinacker:2011wb}, and there are plenty 
of more sophisticated compactifications  
\cite{comp-preparation}. Moreover,  we only need 
(near-) Einstein gravity in the 4 non-compact direction, and not on $\cK$.
It remains to be seen if a realistic 4-dimensional gravity can be obtained 
for suitable compactifications.
If so,  this could provide a very appealing theory for gravity which is not only well-suited for 
quantization, but  also protected from  the usual fine-tuning problems.

It should  be clear that this mechanism is completely unavoidable 
on branes of the structure $\cM = \cM^4 \times \cK\subset \R^D$ in matrix models. 
It  implies a long-range gravity-like force on $\cM^4$, which would certainly dominate the 
bulk gravity with its $r^{-8}$ Newton law at long distances. Hence there 
is no need for 10-dimensional compactification, and
the selection of 
the present type of compactification is a well-defined and predictive  question within the 
matrix model.

Finally, we recall that short-range perturbations with 
$\nabla \cJ^2 \neq 0$ are expected in the presence of matter, as discussed in section \ref{sec:e-m-cons}.

\paragraph{Towards 4-dimension gravity.}

Although the above results apply to any $\cM\subset \R^D$, we are mainly interested in the 
low-energy sector on backgrounds of the form
$\cM = \cM^4 \times \cK \subset \R^D$. We should therefore perform an appropriate reduction on $\cK$,
and study the 4-dimensional effective geometry.
This reduction is not  trivial here, because $\cK$ is {\em not}
perpendicular to $\cM^4$. As discussed in \cite{Steinacker:2012ra}, the effective 4-dimensional metric 
with upper (!) indices 
$G^{\mu\nu}$ is  obtained from $G^{ab}$ by dropping the extra coordinates\footnote{Due to the 
flux stabilization mechanism, we can assume here that $r_\cK = const$ as discussed below.},
and averaging over $\cK$ if necessary
\begin{align}
G^{\mu\nu}_{4D} := \int_\cK G^{\mu\nu} \ .
\end{align}
Here we assume that the low-energy physical fields are constant along $\cK$ (for the 
lowest KK modes), which moreover has constant radius as discussed in section \ref{sec:radial}. 
However, the inverse effective 4-dimensional metric 
$G_{\mu\nu}^{4D}$ is in general {\em not} such a simple reduction of $G_{ab}$;
non-compact coordinates are indicated by Greek letters.
We should thus be careful before drawing physical conclusions, but the salient features are expected to survive.
In particular, the term 
\begin{align}
\g^{bd} K_{ac}K_{bd}^\dagger
\end{align}
 is certainly large on $\cK$ but should typically vanish on $\cM^4$,
consistent with the fact that the 4-dimensional geometry
is  flat for the basic solutions found in \cite{Steinacker:2011wb,comp-preparation}, in the absence of matter.
In particular, assuming that  $\cP^{ab;cd}$ is Lorentz-invariant
with respect to the 4-dimensional effective metric\footnote{Lorentz invariance with respect to the full 
metric on $\cM$ is presumably too restrictive, and we do not expect that Einstein gravity is recovered on $\cM^{2n}$.
Moreover only a part of the full tensor $\cP$ is used  after the reduction, so that
even the effective sign is not clear at this point.}
and assuming that the properly reduced equations
\eq{Ricci-gamma-J-NC} have the same form, we would indeed obtain the Einstein equations, possibly with additional 
vacuum contributions due to $g_{\eta\s} \cP^{\mu\eta;\nu\s}$.
The effective gravitational constant is set by the 
scales of compactification and $\L_0$ \cite{Steinacker:2012ra},
\be
G_N \sim  r_\cK^{-2} \L_0^{-4} \ .
\ee
Although this requires several assumptions about the background  (most importantly 
effective Lorentz invariance of $\cP$ ), the message is that an effective gravity  similar to Einstein gravity
can arise
from  compactified branes $\cM^4\times \cK \subset \R^D$ in matrix models,
 without an Einstein-Hilbert action.
The physical meaning of the  additional term $g_{\eta\s} \cP^{\mu\eta;\nu\s}$ remains to be clarified. 
It might contribute an additional coupling to $T_{\mu\nu}$, and it will probably 
contribute constant tensors such as $\g^{\mu\nu}$ or $(\g g \g)^{\mu\nu}$. Furthermore, 
harmonic contributions from $\cM^4$ may also play a role, cf. \cite{Steinacker:2009mp}. It is tempting to speculate that 
these modifications of the Einstein equations might manifest themselves as dark matter and/or energy.

\subsection{Energy-momentum conservation}
\label{sec:e-m-cons}

To understand better possible deviations from $\nabla \cJ^2 = 0$, we study the tangential 
degrees of freedom in more detail. These are conveniently captured by the matrix conservation law \cite{Steinacker:2008ri}
\begin{align}
 0 &=  -i[X_B,\cT^{AB}] \sim \{x_B,\cT^{AB}\}
\end{align}
where $\cT^{AB}$ is the ''matrix`` energy-momentum tensor. 
Dropping the contributions of the spinorial (fermionic) matrices $\Psi$,
it is given explicitly by
\be
 \cT^{AB} = \frac 12 ([X^A,X^C][X^B,X_{C}]  + (A \leftrightarrow B))  - \frac 14 \eta^{AB} [X^C,X^D][X_C,X_D] .
\label{e-m-tens}
\ee
We can split this tensor into geometrical and matter content,
\begin{align}
\cT^{AB} &= \cT^{AB}_{\rm geom}  + \cT^{AB}_{\rm mat}  \
  = \del_a x^A \del_b x^B  \theta^{aa'}\theta^{bb'}\Big(T_{a'b'}^{\rm geom} +  e^\s \L_0^{-4}  T_{a'b'}^{\rm mat}\Big), \nn\\
T_{ab}^{\rm geom} &= - g_{ab}  + \frac 14 \g_{ab} (\g^{cd}g_{cd}) 
\end{align}
and the nonabelian component is essentially the usual energy-momentum tensor, at least for $\theta= const$.
Therefore
\begin{align}
  \{x_B,\cT^{AB}_{\rm geom}\} =-\{x_B,\cT^{AB}_{\rm mat}\} 
\end{align}
 describes the back-reaction of matter to the Poisson structure.
To understand this, we observe  
\begin{align}
\{x_B,\cT^{AB}\} &= \theta^{cd}\del_c x^B \del_d (\cT^{AD})  \eta_{BD} \nn\\
&=  \theta^{cd} \del_d (\del_a x^A\del_c x^B \del_b x^D \theta^{aa'}\theta^{bb'}  T_{a'b'} )  \eta_{BD} \nn\\
&=  \theta^{cd} \del_d ( g_{cb} \theta^{bb'} T_{a'b'} \theta^{aa'}\del_a x^A) 
\label{cons-law-explicit-1}
\end{align}
in any local coordinates.
For the geometric contribution, this can be written as
\begin{align}
\{x_B, \cT^{AB}_{\rm geom}\}  
&=  \theta^{cd} \del_d (\cJ^{-2} - \frac 14 (\tr \cJ^{-2}) \d)^{\ a}_{c}\del_a x^A  \nn\\
&=  \theta^{cd} (\nabla_d \cJ^{-2 a}_{c}  - \frac 14 \del_d (\tr \cJ^{-2})\d_c^a) \del_a x^A 
\label{cons-law-explicit}
\end{align}
(for any torsion-free $\nabla)$
noting that the transversal contribution vanish, in particular
\begin{align}
 \g^{db} g_{a'b} \theta^{aa'}\nabla_d \del_a x^A  
 = (\theta^{dd'} g_{d'b'} \theta^{bb'}g_{ba} \theta^{ac})\nabla_d \del_c x^A  = 0 
\end{align}
since $\theta g \theta g \theta$ is anti-symmetric.
Thus \eq{cons-law-explicit} is purely tangential.
For the matter contribution, we can proceed as follows
\begin{align}
  \L_0^{4} \{x_B,\cT^{AB}_{\rm matter}\} &= \frac{e^\s}{\sqrt{G}} \del_d (\sqrt{G} G^{db}  T_{ab} \theta^{a'a}\del_{a'} x^A) \nn\\
 &= e^\s \Big(G^{db}(x)\,\nabla_d[G] T_{ba} - \frac 12 \del_{a}  G^{db} T_{b d} \Big) \theta^{a'a}\del_{a'} x^A
 + e^\s G^{db}  T_{ab}  \del_d (\theta^{a'a}\del_{a'} x^A) \nn\\
&\stackrel{\rm cons}{=} e^\s \Big( - \frac 12  T_{b d}\nabla_{a}[g]  G^{db} \cJ^{-1 a}_e\
 + G^{db}  T_{ba} \nabla_d[g] \cJ^{-1 a}_e \Big) g^{ea'} \del_{a'} x^A
 + (...)\nabla[g]\del x^A    \nn
\end{align}
using the identity  \cite{Steinacker:2008ri}
 \begin{align}
  \rho \nabla_b \theta^{bc} &= \theta^{cb}\del_b \rho, \qquad \rho =  \sqrt{|\theta^{-1}|}
\label{del-theta-id}
 \end{align}
 and  \eq{cov-cons-law} in the appendix.
The first two lines hold in any coordinates, and energy-momentum conservation
$\nabla^b[G] T_{ba} = 0$ was assumed in the last line. 
We  choose normal embedding coordinates such that $\del\del x^A = \nabla[g]\del x^A$ is 
in the normal bundle, and together with \eq{cons-law-explicit}  the tangential components give
\begin{align}
 \theta^{cd} \nabla_d \cJ^{-2 a}_{c}  - \frac 14 \theta^{ad} \del_d (\tr \cJ^{-2})\,
 = - \frac{e^\s}2  \L_0^{-4} T_{b d}\nabla_c[g]  G^{db} \cJ^{-1 c}_e\,g^{ae}
 + e^\s  \L_0^{-4}  G^{db} T_{bc} \nabla_d[g] \cJ^{-1 c}_e\,g^{ae} .
\end{align}
Therefore any vacuum geometry with $\nabla \cJ^2 = 0$ is a solution. 
Short-range perturbations of  $\nabla\cJ^2$ are expected 
in the presence of matter, which do not significantly contribute to 
gravity at long distances.
To see this, it is better to use the fundamental degrees of freedom 
given by the Poisson structure and the embedding. 
Using the identity \eq{t-geom-cons-g}, the same  conservation law can be written
as follows
\begin{align}
  \g^{da}\nabla_a[g]\theta^{-1}_{bd} \, 
 =  \frac{e^\s}2  \L_0^{-4} T_{b d}\nabla_c[g]  G^{db} \theta^{ca}\g_{ba}\, 
 - e^\s \L_0^{-4} G^{db}   T_{bc} \nabla_d[g]  \theta^{ca}\g_{ba}\, .
\end{align}
Since this has the structure of Maxwell equations, the perturbations of $\theta^{-1}_{bd}$ due to matter
decay at least as $\frac 1{r^2}$, and therefore do not contribute to gravity at long distances.
This  is consistent\footnote{The assumption
$\Gamma^a=0$ in  \cite{Steinacker:2012ra} amounts to $\{x_B, \cT^{AB}_{\rm geom}\} =0$ via \eq{Tgeom-cons-Gamma},
and  therefore follows from
 $\nabla \cJ^2 = 0$.} with the equation  (3.6) in \cite{Steinacker:2012ra},
 which was obtained directly from the action.

\subsection{Radial equation of motion and flux stabilization}
\label{sec:radial}

The equation of motion for the radial mode $r^2(x) = x^A x_A$ can be derived using the  identity
\eq{deldelr-id}, which gives
\begin{align}
\frac 12  \Box r^2    &= r \Box r =  \g^{ab }(g_{ab} + K^0_{ab}) 
=   \g^{ab }g_{ab} -  \L_0^{-4} T_{cd}\,\Pi^{cd}_{a'b'}\, \theta^{a'a}\theta^{b'b} \, K_{ab}^0
\end{align}
Since we argued or assumed that $\nabla \cJ^2 = 0$
to a very good approximation, it follows that $ (\g g) = - \tr \cJ^{-2} = const$.
This vanishes if and only if the action is invariant under $x^A \to \a x^A$, and one may expect that
this is preferred upon quantization.

Now consider the case of compactified extra dimensions $M^4 \times \cK \subset \R^{10}$ where 
$\cK \subset \R^6$ is compact.
We can locally write $\R^{10} = \R^4 \times \R^6$ with $x^A = (x^\mu,y^i)$ such that the radius is 
$r_\cK^2 = y_i y^i$, and use the 4 non-compact matrices $x^\mu$
as part of the local coordinates $\xi^a = (x^\mu,\xi^i)$. 
Then the equation of motion for $r_\cK$ can be obtained as follows:
\begin{align}
 \Box r^2  &= (\g^{\mu\nu}\del_\mu\del_\nu) (x^\rho x^\sigma \eta_{\rho\sigma})  + \Box r_\cK^2 
 = 2 \g^{\mu\nu}\eta_{\mu\nu}  + \Box r_\cK^2 
 \label{Box-r-rK}
\end{align}
in NEC.
Together with the above we obtain 
\begin{align}
\frac 12  \Box r_\cK^2 &= 2 \g^{i\mu}g_{i\mu} + \g^{ij} g_{ij} \nn\\
&= 2 \g^{i\mu}g_{i\mu} + g_{ij} \theta^{ii'}\theta^{jj'} g_{i'j'} = f(r_\cK) 
\end{align}
in vacuum.
This is a polynomial in $r_\cK$ via $g_{ij} \sim r_\cK^2$ .
If the flux $\theta^{ij}$ on $\cK$ does not vanish,
then the rhs contains quadratic and quartic terms in $r_\cK$, and will vanish
for a certain radius $r_0$ for suitable $\theta^{\mu i}$. The radial perturbations of the compactification $\cK$ 
are then in general stabilized by the flux and (very) massive, so that
we can safely set $r_\cK = const$ at low energies. 
This is the flux stabilization mechanism in the present context.

\section{Perturbations of the  geometry}
\label{sec:perturbations}

Consider a  perturbation  
\begin{align}
x^A \ \to \ x^A + \d x^A 
\end{align}
of some background brane $\cM^{2n}\subset \R^D$, defined in terms of matrices $X^A\sim x^A$ as above.
We can certainly describe the most general such deformations in the form
\begin{align} 
 \d x^A = -\sum_{\a\neq 0} \e_\a (\l^\a x)^A \, + \, \d r\,  x^A
\label{variation-symm} 
\end{align} 
where $\e_\a = \e_\a(x)$ and $\e_0 \equiv \d r = \d r(x)$ are arbitrary functions.
This is of course an over-parametrization.  
The corresponding metric perturbation can be written in terms of the currents as 
\begin{align}
 \d g_{ab} &= -\del_a x \del_b (\e_\a(x) \l^\a x)\  + \del_a x \del_b (\d r(x) x)\ + (a \leftrightarrow b) \nn\\
 &=  J_a^{\a} \del_b \e_\a +  J_b^{\a} \del_a \e_\a +  \ 2\e_0\, g_{ab}  
\label{d-g-full}
\end{align}
since $\l^{\a\neq 0}$ is anti-symmetric.
To clarify the relation with the approach in \cite{Steinacker:2012ra}, 
we can then rewrite this as
\begin{align}
 \d g_{ab} &= \nabla_b (\e^\a  J_a^{\a}) - \e^\a  \nabla_b J_a^{\a} + \e_0\, g_{ab}   + (a \leftrightarrow b)
\label{d-g-full-2} \\
&= -2\e_\a K^\a_{ab}  + \nabla_a V^\e_b + \nabla_b V^\e_a +  \ 2\e_0\, g_{ab}  
\end{align}
where $\nabla = \nabla[g]$.
Then the vector fields
\begin{align}
V^\e_b = \e^\a J_b^{\a}
\end{align}
 encode the tangential perturbations, while the extrinsic curvature  of $\cM \subset \R^{D}$ 
leads to linearized metric perturbations $-2\e_\a K^\a_{ab}$ due to transversal 
brane perturbations.

\subsection{Current conservation and matter}
\label{sec:cons-current-matter}

In the presence of matter, the $SO(D)$ rotations also act on the fermions and gauge fields.
Rather than trying to derive the generalized currents, we want to incorporate matter as source term
for the  conservation law of the geometrical current \eq{current-cons-matr}.
We therefore need the variation of the matter action under the local
perturbations  \eq{variation-symm} acting only on the geometry defined by the $U(1)$ sector of the 
matrices $X^A \sim x^A$, in the presence of fixed matter fields resp. matrices (on-shell).
Restricting ourselves to the semi-classical case, matter couples to the background as usual 
via the effective metric $G$. Therefore  the variation of the action under these geometrical
 $SO(D)$ rotations is simply obtained by the 
energy-momentum tensor $T_{ab}$ of matter coupled to $\d G_{ab}$. 
We choose to work in Darboux coordinates where $\theta^{ab} = const$ is fixed, 
which is always possible\footnote{From the point of view of noncommutative gauge theory on $\R^{2n}_\theta$, this means that 
all matter fields and $SU(n)$-valued fields are fixed, and only the trace-$U(1)$ scalar fields are perturbed.
The latter are interpreted as perturbations of the embedding metric $\d g_{ab}$. }. 
Then the variation of the effective metric \eq{eff-metric} takes the form
\begin{align}
 \d G^{ab} &=  e^{-\s}\Pi_{cd}^{ab}\,\theta^{cc'}\theta^{dd'}  \d g_{c'd'} 
\end{align}
where
\begin{align}
 \Pi^{cd}_{ab} = \d^{cd}_{ab} - \frac{\g_{ab}\g^{cd}}{2(n-1)}
\end{align}
Then
\begin{align}
\d S_{\rm YM} + \d S_{\rm matter} &= \frac{1}{2(2\pi)^n} \int d^{2n}x\, 
 \Big(\L_0^4\sqrt{\theta^{-1}} \g^{ab}\d g_{ab}
 +\sqrt{G} \, T_{ab}  \,  \d G^{ab}\Big) \nn\\
&= \frac{1}{2(2\pi)^n} \int d^{2n}x\, \sqrt{\theta^{-1}} \Big(\L_0^4\g^{ab} 
 +\, T_{cd}\, \Pi_{a'b'}^{cd}\, \theta^{a'a}\theta^{b'b}  \Big)\d g_{ab} 
\label{dS-full}
\end{align}
where $\d g_{ab}$ is given by  \eq{d-g-full}. Upon partial integration,
we obtain the current conservation law in the presence of matter 
\begin{align}
  \del_a(\g^{ab} J_b^\a) &= - \L_0^{-4}\del_a\big(T_{cd}\, \Pi_{a'b'}^{cd}\, \theta^{a'a}\theta^{b'b}  J_b^\a\big), \qquad \a\neq 0 \nn\\
   \del_a(\g^{ab} J_b^0)   &= -  \L_0^{-4}T_{cd}\, \Pi_{a'b'}^{cd}\, \theta^{a'a}\theta^{b'b} K^0_{ab} + (\g^{ab}g_{ab})
 \end{align}
 The second equation follows recalling that $\del_a J^0_b = g_{ab} + K^0_{ab}$ \eq{J-T-K-id}, 
 and setting $J^0|_p \sim\del r|_P = 0$ after a suitable translation.
The lhs can be written covariantly using \eq{Gamma-id}, and  we obtain 
\begin{align}
 e^\s\nabla^a[G] J_a^\a &= \g^{ab} K_{ab}^\a \ + \cO(J^\a) 
  = -  \L_0^{-4} T_{cd}\,\Pi^{cd}_{a'b'}\, \theta^{a'a}\theta^{b'b} \,K_{ab}^\a \ + \cO(J^\a) , \qquad \a\neq 0 \nn\\
 e^\s\nabla^a[G] J_a^0 - (\g^{ab}g_{ab}) &= \g^{ab} K^0_{ab}|_p  = 
  -  \L_0^{-4} T_{cd}\,\Pi^{cd}_{a'b'}\, \theta^{a'a}\theta^{b'b} \, K_{ab}^0 
\label{cons-law-matter}
\end{align}
Note that $\cO(J^\a), \ \a\neq 0$  drops out  from the equation  \eq{Ricci-gamma-J-NC} for the Ricci tensor
because it is tangential, while  $K_{cd}^\a$ is transversal.
The basic mechanism can now be seen by 
observing that  current conservation \eq{currents-semiclass}
\begin{align}
 e^\s \nabla^a[G] J_a^\a &=   \g^{ab} K^\a_{ab} =  x \l^\a \Box_G x, \quad \a\neq 0, \nn\\
  e^\s\nabla^a[G] J_a^0- \g^{ab}g_{ab} &=  \g^{ab} K^0_{ab}|_p  = x  \Box_G x
\end{align}
measures the deviation from harmonicity of the  embedding, which couples via $K_{ab}^\a$
to the energy-momentum tensor, and contributes to ${\rm Ric}^{ab}[\g]$. This is the same mechanism as in \cite{Steinacker:2012ra}.

\section{Conclusion}

In this paper, a formalism for 
computing the effective curvature of branes in the matrix model is developed.
This is done by describing the geometry in terms of an over-complete frame,  based on the 
currents associated with the global $SO(D)$ symmetry of the model.
One result is that 
the effective Ricci tensor has  contributions which couple linearly to the energy-momentum tensor.
However the coupling is not  direct as in general relativity, but
somewhat implicit via a coupling tensor $\cP$ which depends
on the Poisson tensor and the extrinsic curvature of the brane embedding $\cM \subset \R^D$. 
An extra term may lead to  vacuum solutions which are not Ricci flat. 
This mechanism is particularly significant for compactified branes
$\cM = \cM^4 \times \cK \subset \R^D$, where the coupling $\cP$ is always non-vanishing.
While the detailed physical consequences depend on the compactification and
remain to be clarified, the mechanism clearly leads to a gravity-like long-range force 
on compactified brane solutions in matrix models, 
which is not based on the Einstein-Hilbert action.
The relation with global symmetries and with non-commutative gauge theory 
make this mechanism for gravity very attractive for quantization, notably for the maximally
supersymmetric IKKT model.

Having confirmed the basic mechanism observed in \cite{Steinacker:2012ra}, 
the tools provided here should allow a more detailed study of the resulting gravity theory. In particular, 
the additional terms  in the geometric equation \eq{Ricci-gamma-J-NC} due to $\cP$ 
need to be understood,  the response of $\cP$ to matter must be clarified, and suitable compactifications 
must be found.
If the resulting gravity turns out to be viable, this would have far-reaching implications.
Since  target space does not need to be compactified, the
vast landscape of 10-dimensional compactifications and its inherent lack of predictivity could be discarded.
It suffices instead to consider lower-dimensional brane compactifications of  type $\cM^4 \times \cK \subset \R^{10}$,
which  may also provide the additional structure required for particle physics \cite{Chatzistavrakidis:2011gs}.
Note that there is no contradiction with string theory: the 10-dimensional bulk gravity does indeed arise
in a holographic sense. However, bulk gravity is not the dominant mechanism on branes of type 
$\cM^4 \times \cK \subset \R^{10}$ with $B$ -field,
since the present mechanism leads to a
 4-dimensional effective gravity, which is clearly dominant for long distances.
Note also that in the matrix model there are a priori no propagating degrees of freedom in the bulk,
so that we expect no problem with energy leaking off the brane.
This is certainly sufficient motivation for more detailed studies.

\paragraph{Acknowledgments.}

Part of this work evolved during an extended visit at the high-energy physics group at CUNY. 
The hospitality and useful discussions in particular with  A. Polychronakos,
D. Kabat and D. Karabali and P. Nair are gratefully acknowledged. 
I also thank J. Zahn and P. Schreivogel for discussions.
This work  was supported in part by the Austrian Science Fund (FWF) under the contract
P24713, and in part by a CCNY/Lehman CUNY collaborative grant.

\startappendix

\Appendix{Conserved currents}
\label{sec:currents}

We want to derive the conservation law corresponding to the $SO(D)$ symmetry, 
which acts as
\begin{align}
\d X^A = \l^A_B X^B \, 
\end{align}
for some $\l \in \mso(D)$.
Consider the  corresponding ''local`` transformation
\begin{align}
 \d_\e X^A =  \frac 12\l^A_B \{\e(X),X^B\} \, .
\end{align}
The corresponding variation of the action is
\begin{align}
 \d S &= Tr\, \d X_A\Box X^A =  \frac 12 Tr\, \l_{AB} \{\e(X),X^B\}\, \Box X^A  \nn\\
 &=  \frac 12 Tr\,\e(X) \l_{AB} \{X^B,[X_C,[X^C,X^A]] \} \ .
\end{align}
Using the identity
\begin{align}
 \{A,[B,C]\} = [B,\{A,C\}] - \{C,[B,A]\}
\end{align}
this becomes
\begin{align}
 \d S&=  \frac 12 Tr\,\e(X) \l_{AB} \big([X_C,\{ X^B,[X^C,X^A]\}] - \{[X^C,X^A],[X_C,X^B ]\} \big) \nn\\
 &=  \frac 12 Tr\,\e(X) \l_{AB} \big([X_C, \{ X^B,[X^C,X^A]\}])
\end{align}
as the second term vanishes identically; this reflects the invariance under rigid transformations.
This vanishes on-shell for any $\e(X)$, and we obtain the conservation law 
\begin{align}
 [X_A, \tilde J^A] = 0, \qquad
 \tilde J^C &=  \frac 12 \{\l_{AB} X^A,[X^C, X^B]\} \sim i \theta^{ab} \del_a X^C J_b, \nn\\
 J_b &=   \l_{AB} x^A\del_b x^B .
\end{align}
This can also be verified directly using the equations of motion \eq{eom-IKKT}.
Note that $\tilde J^A$ is  a tangential vector field on $\cM \subset \R^{10}$.
In the semi-classical limit, the conservation law amounts to
\begin{align}
0 &=\theta^{bc}\del_b X_A \del_c (\theta^{ae} \del_a X^A J_e) = 
 \theta^{bc}\del_c (g_{ab}\theta^{a e} J_e) \nn\\
 &= \g^{ce} \del_c J_e + \rho^{-1} \del_c(\rho\g^{ce}) J_e 
 = e^\s\big(G^{ce} \del_c J_e -\Gamma^e[G] )J_e \nn\\
 &=  e^\s \nabla^e[G] J_e 
\label{current-cons-g}
\end{align}
 using the identity \eq{del-theta-id} for $\rho = \sqrt{|\theta^{-1}|}$, and
recalling that
\begin{align}
- \Gamma^a[G] &=\frac 1{\sqrt{|G|}}\del_b(\sqrt{|G|}G^{ab})  
 = \rho^{-1} e^{-\s}\del_b(\rho\g^{ab}) \ .
 \label{Gamma-id}
\end{align}
This is  the usual covariant conservation law, once again confirming $G$ as the relevant metric. 
In Darboux coordinates, this conservation law reduces to
\begin{align}
e^\s \nabla^c[G] J_c \equiv  \del_c (\g^{ce} J_e) &= 0 \ .
 \label{cons-current-darboux}
\end{align}

\Appendix{Currents and structure constants}
\label{sec:deriv-currents}

We observe the following identity for $\mso(D)$ (resp. $\mso(1,D-1)$) generators
\begin{align}
 f^\a_{\b\c} \l^\b_{AB} \l^\c_{CD} =  2\eta_{BC} \l^\a_{AD}  + 2\eta_{AD} \l^\a_{BC}
- 2\eta_{BD} \l^\a_{AC} - 2\eta_{AC} \l^\a_{BD}  
\label{f-strange-id}
\end{align}
where $f^\a_{\b\c}$ are the structure constants of $\mso(D)$.
This can be established using the basis $\l^\a = \begin{pmatrix}
                                              0 & 1 \\ -1 & 0
                                             \end{pmatrix}$.
For the currents $J^\a$, this implies
\begin{align}
 f^\a_{\b\c} J^\b_b J^\g_c &= f^\a_{\b\c} (x^A\l^\b_{AB}\del_b x^B) (x^C\l^\c_{CD} \del_c x^D)    \nn\\
 &= \del_b r^2 x^A\l^\a_{AD} \del_c x^D  + \del_c r^2  x^C\l^\a_{BC} \del_b x^B
 - 2 r^2 \del_b x^B\l^\a_{BD} \del_c x^D   \nn\\
 &= \del_b r^2 J_c^\a  - \del_c r^2  J_b^\a - 2r^2 T_{bc}^\a   \qquad (\a\neq 0)
\label{strange-id-T-app}
\end{align}
and therefore
\begin{align}
r^{-2} f_{\a\b\c} J^\a_a J^\b_b J^\g_c  &=  \del_b r^2 g_{ac} - \del_c r^2 g_{ab} - 2\sum_{\a\neq 0} J^\a_a T_{bc}^\b \k_{\a\b}\nn\\
  &=  2r(\del_b r g_{ac} - \del_c r g_{ab} -  \del_b r g_{ac} + \del_c r g_{ab} 
  - \del_a r g_{bc} ) +2r\del_a r g_{bc} \nn\\
  &= 0  \nn\\
 f^\a_{\b\c} J^\b J^\g &=  2 d r^2 J^\a  - 2 r^2  d J^\a \ .
\label{strange-id-app}
\end{align}
The first identity can also be seen in NEC, and 
the last identity also holds for $\a=0$, where both sides vanish.
Combining these, \eq{strange-id-app}  implies
\begin{align}
d\theta^\a P^{\a'\b} \k_{\a\a'} = \frac{dr}{r} \theta^\b = \frac 1r\, \theta^0\, \theta^\b 
\end{align}
which using $\theta^\a \theta^\b \k_{\a\b} = 0$ gives \eq{dtheta-result}.

\Appendix{Radial curvature}
\label{sec:curv-rad}

We compute the curvature contribution due to $\omega$. 
After a suitable translation
(or working in NEC)  all first-order derivatives $\del r$ such as $\omega$ can be dropped, 
but we must keep the second derivatives:
\begin{align}
 d(r\omega^{\a\b}) &=  d\theta^\a P^{(r)\b} - \theta^\a d P^{(r)\b}  - d P^{(r)\a} \theta^\b - P^{(r)\a} d\theta^\b \nn\\
r (P d\omega P)^{\a\b} 
  &= (P d\theta)^\a P^{(r)\b} - \theta^\a (P d P)^{\b(r)}  - (Pd P)^{\a (r)} \theta^\b - P^{(r)\a} (P d\theta)^\b \nn\\
  &= - \theta^\a (P d P)^{\b(r)}  - (Pd P)^{\a (r)} \theta^\b 
\end{align}
since $P d\theta = -\theta \omega$ can be dropped.
Using $P^{(r)\a} = \del_a r g^{ab} \theta_b^\a$ we get
\begin{align}
 (P d P)^{\b(r)} = \del_f \del_d r g^{db} \theta_b^\b\, dx^f \ .
\end{align}
Therefore 
\begin{align}
 R[\omega]_{ac} &= \theta_a ( d\omega + \omega\omega) \theta_c = \theta_a  d\omega\, \theta_c \nn\\
 &= -  \theta_a \theta_e^\a r^{-1}\nabla_f \del_d r g^{db} \theta_b^\b  \theta_c \, dx^e dx^f
  -  \theta_a r^{-1}\nabla_f \del_d r g^{db} \theta_b^\a  \theta_c \theta_e^\b\,  dx^f dx^e\nn\\
 &=  r^{-2}\big( - g_{ae} \nabla_f J_c^0   \, +  g_{ce} \nabla_f J^0_a  \,\big)\,  dx^e dx^f \nn\\
  &= R[\omega]_{acef} dx^e dx^f 
  \label{omega-curv-explicit} 
\end{align}
where $\nabla = \nabla[g]$,
using \eq{deldelr-id} and recalling that $\nabla_a J^0_{b} = \frac 12 \nabla_a\del_b r^2$.
Furthermore, we need the contraction 
\begin{align}
 r^2 g^{ae}\, \cR_{ac;ef}[\omega] &=  2(1-n) \nabla_f J_c^0   \, -  g_{cf} (g^{ae}\nabla_e J^0_a) \ .
  \label{R-omega-contract}
\end{align}

\Appendix{Curvature tensor for special geometries}
\label{sec:spec-geom-curv}

We note the following identities
\begin{align}
P d\L^{-1} P_N &= P (\one - \L^{-1}) d P_N \nn\\
 P_N d \L P &= d P_N (\one -\L )P
\end{align}
which follow from $\L P_N = P_N$.
Using this and
assuming the condition $\nabla Q =0$ such that  $B^{(\a)}_\b = 0$, we obtain using \eq{A-gamma}
\begin{align}
  \Theta_a A A \Theta_b^\dagger  &= \Theta_a  \L^{-1} d \L P \L^{-1} d \L \Theta_b^\dagger 
   = - \Theta_a d \L^{-1} P d \L  \Theta_b^\dagger \nn\\
   &= - \Theta_a d \L^{-1} d \L \Theta_b^\dagger + \Theta_a d \L^{-1}P_N d \L \Theta_b^\dagger  \nn\\
  &=  - \Theta_a d \L^{-1} d \L \Theta_b^\dagger 
   +  \Theta_a(\one -\L^{-1})d P_Nd P_N (\one - \L) \Theta_b^\dagger .
\end{align}
To evaluate
\begin{align}
 \Theta_a dA\Theta_b^\dagger  &= \Theta_a\big(dP \L^{-1} d \L P + P d\L^{-1} d \L P - P \L^{-1} d \L dP  \big) \Theta_b^\dagger 
\end{align}
we observe that  $P dP_N P = 0$, which implies the following useful identity
\begin{align}
 \theta dP &= -\theta d P_N =  -\theta d P_N P_N  \nn\\
 dP \theta^\dagger &= - dP_N \theta^\dagger = - P_N dP_N \theta^\dagger .
\end{align}
This gives
\begin{align}
- \Theta_a P \L^{-1} d \L dP \Theta_b^\dagger &=  \theta_a d \L P_N dP_N  \Theta_b^\dagger 
 = \theta_a(\one - \L ) d P_N dP_N  \Theta_b^\dagger 
\end{align}
as well as
\begin{align}
 \Theta_a dP \L^{-1} d \L P \Theta_b^\dagger &= -\Theta_a dP_N P_N  d \L \Theta_b^\dagger 
 = -\Theta_a dP_N d P_N (\one -\L )\Theta_b^\dagger \ .
\end{align}
Then the metric curvature tensor is obtained using \eq{R-gamma-expand}
\begin{align}
 R_{ab}[\g] &= \theta_a \L dP_N dP_N \L^\dagger \theta_b^\dagger + \Theta_a (dA + A A)\Theta_b^\dagger  \nn\\
 &=  \Theta_a dP_N dP_N  \Theta_b^\dagger 
   +  \Theta_a(\one -\L^{-1})d P_Nd P_N (\one - \L) \Theta_b^\dagger \nn\\
 & \quad   + \theta_a(\one - \L )  d P_N dP_N  \Theta_b^\dagger 
  -\Theta_a dP_N d P_N (\one -\L )\Theta_b^\dagger 
  \ + \Theta_a \L^{-1} d\omega \L \Theta_b^\dagger  \nn\\
 &= \theta_a (d P_Nd P_N +d\omega) \L \Theta_b^\dagger \nn\\
 &=  R_{ab}[g] +  \theta_a d P_Nd P_N  Q \theta_b^\dagger 
 \label{R-gamma-appendix}
\end{align}
recalling that $\L\L^\dagger = Q + \one$, and dropping $\omega \sim \del r$ after a suitable translation (or in NEC).

\Appendix{Covariance of conservation laws}

Consider
\bea
 G^{ca}(x)\,\nabla_c[G]\, T_{ab } &=&
 G^{ca}(x)\,\(\partial_c T_{ab } 
-  \Gamma_{ca}^d T_{d b }
-  \Gamma_{cb }^d T_{a d}\) \nn\\
&=&  G^{ca }\,\partial_c T_{a b } 
-  \Gamma^d T_{d b }
-  G^{ca }\, \Gamma_{cb }^d T_{a d} 
\eea
where 
\be
 \Gamma^c =  G^{ab} \Gamma^c_{ab} = 
- \frac 1{\sqrt{ G}} \partial_d( G^{c d} \sqrt{ G}) \, .
\label{Gamma-def}
\ee
we can write 
\bea
 G^{ca}\, \Gamma_{cb }^d  T_{a  d }
= \frac 12  G^{ca}\,  G^{\rho  d } T_{a  d }
\(\partial_c  G_{e b }+ \partial_b   G_{e c} - \partial_e  G_{cb } \)
 = \frac 12 T^{ce}\partial_b   G_{e c}
\eea
where $T^{c e } =  G^{ca}  G^{e  d }\, T_{a  d }$.
Therefore
\bea
 G^{ca}(x)\,\nabla_c[G]\, T_{ab } 
&=& G^{ca }\,\partial_c T_{a b } 
- \frac 12  T^{c\rho}\partial_b   G_{\rho c} -  \Gamma^\rho T_{\rho b }  \nn\\
&=&  \frac 1{\sqrt{ G}} \partial_c\big( G^{c a} \sqrt{ G} \, T_{ab }\big)
+  \frac 12 \partial_b   G^{ca} T_{a c} ,
\label{cov-cons-law}
\eea
where the rhs is valid for any connection.

Finally, we recall the  identity (see (2.51), (2.53) in \cite{Blaschke:2010qj})
\begin{align}
 \{X_B,\cT^{AB}_{\rm geom}\} = e^\s \Box_G x^B\del_a x_B \theta^{ae}\del_e x^A
 \label{Tgeom-cons-Gamma}
\end{align}
and note that
\begin{align}
e^\s \Box_G x^b   
 &= \{x^A,\{x_A,x^b\}\}  = \theta^{ac}\del_a(\theta^{bd}g_{dc}) \nn\\
 &= \theta^{ac}g_{dc}\nabla_a[g]\theta^{bd}
 =  - \g^{da}\nabla_a[g]\theta^{-1}_{cd} \, \theta^{bc} .
\end{align}
Combining these relations gives
\begin{align}
 \{X_B,\cT^{AB}_{\rm geom}\} 
&= - \g^{da}\nabla_a[g]\theta^{-1}_{bd} \, \g^{be}\, \del_e x^A .
\label{t-geom-cons-g}
\end{align}


\end{document}